%% file: main.tex
\documentclass[sigconf]{acmart}

\usepackage{xcolor}

\def\BibTeX{{\rm B\kern-.05em{\sc i\kern-.025em b}\kern-.08emT\kern-.1667em\lower.7ex\hbox{E}\kern-.125emX}}
    
\copyrightyear{2019}
\acmYear{2019}
\setcopyright{rightsretained}
\acmConference[AsiaCCS '19] {ACM Asia Conference on Computer and Communications Security}{July 9--12, 2019}{Auckland, New Zealand}
\acmBooktitle{ACM Asia Conference on Computer and Communications Security (AsiaCCS '19), July 9--12, 2019, Auckland, New Zealand}
\acmPrice{}
\acmDOI{10.1145/3321705.3329813}
\acmISBN{978-1-4503-6752-3/19/07}

\usepackage{listings}
\usepackage{paralist}

\begin{document}

%
\title{Misbinding Attacks on Secure Device Pairing and Bootstrapping}

%

\author{Mohit Sethi$^{\ast\dagger}$, Aleksi Peltonen$^{\dagger}$, Tuomas Aura$^{\dagger}$}
\affiliation{%
  \institution{$^{\ast}$NomadicLab, Ericsson Research, Finland}
}
\affiliation{%
\institution{$^{\dagger}$Aalto University, Finland}
}
\email{{mohit.sethi, aleksi.peltonen, tuomas.aura}@aalto.fi}



\renewcommand{\authors}{Mohit Sethi, Aleksi Peltonen and Tuomas Aura}

%
\renewcommand{\shortauthors}{Mohit Sethi, Aleksi Peltonen and Tuomas Aura}

%
\input{abstract.tex}

%
%


%

%
\maketitle

\input{introduction2}
\input{background2}
\input{inpairing}
\input{bluetooth}

\input{eapnoob2}
\input{proverif}
\input{mitigation2}

\input{discussion}

\input{conclusion}

\input{acknowledgment}

%
\bibliographystyle{ACM-Reference-Format}
\bibliography{mohitbib,aleksibib}

%

\end{document}

%% file: abstract.tex

\begin{abstract}
In identity misbinding attacks against authenticated key-exchange protocols, a legitimate but compromised participant manipulates the honest parties so that the victim becomes unknowingly associated with a third party. These attacks are well known, and resistance to misbinding is considered a critical requirement for security protocols on the Internet. In the context of device pairing, on the other hand, the attack has received little attention outside the trusted-computing community. This paper points out that most device pairing protocols are vulnerable to misbinding. Device pairing protocols are characterized by lack of a-priory information, such as identifiers and cryptographic roots of trust, about the other endpoint. Therefore, the devices in pairing protocols need to be identified by the user's physical access to them. As case studies for demonstrating the misbinding vulnerability, we use Bluetooth and a protocol that registers new IoT devices to authentication servers on wireless networks. We have implemented the attacks. We also show how the attacks can be found in formal models of the protocols with carefully formulated correspondence assertions. The formal analysis yields a new type of double misbinding attack. While pairing protocols have been extensively modelled and analyzed, misbinding seems to be an aspect that has not previously received sufficient attention. Finally, we discuss potential ways to mitigate the threat and its significance to security of pairing protocols.
\end{abstract}

%% file: introduction2.tex

\section{Introduction}

Secure device pairing is a process that bootstraps secure communication between two physical devices. It is a type of authenticated key-exchange, but with the special characteristic that the endpoints are physical devices which the user can see or touch directly. Unlike most security protocols, secure device pairing does not require pre-established cryptographic credentials or security infrastructure. Instead, the user acts as an out-of-band communications channel or a trusted party that provides the initial security. 

The focus of this paper is on \textit{identity-misbinding}~\cite{krawczyk2003sigma} or \textit{unknown-key-share} attacks~\cite{blake1999unknown} where the wrong endpoints are paired with each other. These attacks depend on one of the user's devices being compromised, and they do not violate the basic secrecy goals. Nevertheless, such vulnerabilities have been considered unacceptable and avoidable in network security protocols. Our main message is that most device-pairing protocols are vulnerable to the misbinding attacks. As we will argue, the vulnerability is not caused by technical errors in the protocol design; rather, it arises from the lack of verifiable identifiers in situations where the endpoint identity is defined by the user's physical access to the device.

This paper is not intended to sound alarm but rather to bring clarity and understanding to a previously ignored question about device authentication. Our contributions are the following: (i) bringing attention to identity-misbinding vulnerabilities in device-pairing and bootstrapping protocols, (ii) detailed analysis and characterization of the vulnerabilities, (ii) examples of concrete, implemented attacks against Bluetooth Secure Simple Pairing and the proposed EAP-NOOB protocol for registering new devices to a network, (iii) formal specification of the violated security property as a correspondence assertion that takes into account the user intention, and (iv) balanced discussion of the impact of these attacks and potential countermeasures. The significance of our work arises from the wide deployment of the vulnerable pairing protocols in everyday applications. 

The rest of the paper is structured as follows. Section \ref{sec:background} discusses the relevant state of the art in security protocols and attacks. Section~\ref{sec:inpairing} explains the misbinding attack against device-pairing protocols and a similar attack when registering new IoT devices to an authentication server. In Section~\ref{sec:proverif}, we show how to model the attack and the related security properties. We also discover a new variant of the misbinding attack. Section~\ref{sec:mitigation} considers the potential solutions. Section \ref{sec:discussion} discusses the significance of the results, and Section~\ref{sec:conclusion} concludes the paper. 

%% file: background2.tex

\section{Background}
\label{sec:background}

\subsection{Security protocol attacks and correspondence assertions}

The goal of authenticated key exchange is to establish a shared cryptographic key between two or more communication endpoints, which then use the shared key for protecting communication integrity and confidentiality. Authenticated key-exchange protocols should be secure against the so-called \textit{Dolev-Yao attacker}~\cite{dolev1983security}, which is able to spoof, intercept and modify messages in the network in arbitrary ways, except when it lacks the necessary cryptographic keys. The attacker may impersonate one of the communication endpoints or set itself as a \textit{man in the middle} (MitM) between them. Even carefully designed protocols have been found to be vulnerable to \textit{forwarding and interleaving attacks}~\cite{abadi1996prudent}\cite{lowe1995attack}, in which the attacker itself is a legitimate participant in the protocol but can mislead others by cleverly replaying messages. In closed systems, such \textit{insider attacks} could sometimes be tolerated, but in large systems and open networks such as the Internet and the Internet of Things, there always are some malicious ``insiders''. Thus, modern security protocols are required to be immune to these attacks. 

The authentication goals of key-exchange protocols can be defined in terms of matching or agreement between the records made by different endpoints on the protocol execution~\cite{diffie1992authentication,lowe1997hierarchy}. The same goals can be stated as \textit{correspondence assertions}~\cite{woo1993semantic}. These assertions define relations between later and earlier events in the protocol execution. For example, a common assertion is that, if Alice accepts a session key to be used with Bob, both Alice and Bob must have previously declared an intent to create such a session key. This way, we can make global assertions about the events that should or should not take place in a distributed system. \textit{Injective} correspondence further requires that each such declaration of intent can result in at most one accepted session key. The assertions are typically parameterized with all the knowledge of protocol inputs and parameters which should match between the events and endpoints.

An advantage of specifying security properties as correspondence assertions is that, in addition to basic authentication properties, the assertions capture the protocol designer's implicit expectations about its execution and, thus, can help to detect subtle flaws that might otherwise go unnoticed. 

\subsection{Identity misbinding}
\label{sec:identitymisbinding}

\begin{figure}[t]
\includegraphics[scale=0.5]{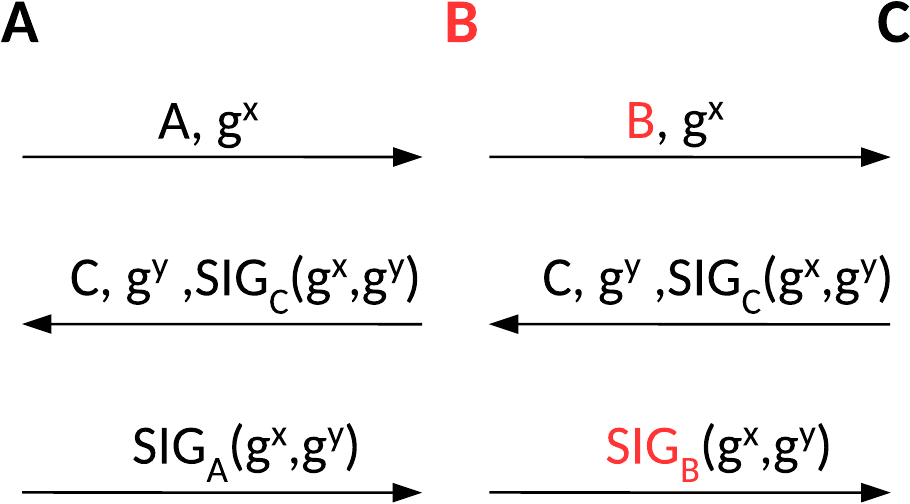}
\caption{Identity misbinding against signed Diffie-Hellman}
\Description{Identity misbinding attack against signed Diffie-Hellman, as described in the literature}
\label{fig:misbinding-original}
\end{figure}

Figure~\ref{fig:misbinding-original} shows an attack on a badly authenticated key exchange. In the figure, the two communication endpoints A and C perform a Diffie-Hellman (DH) key exchange, and the endpoints sign both key shares in order to reach agreement on them. However, a man-in-the-middle attacker B is located between the endpoints and manipulates the messages. In messages travelling from A to C, it replaces A's identifier and signature with its own. This leads to an inconsistency in the states of A and C: A correctly thinks that it shares the session key ${g^{xy}}$ with C, but C has the non-matching belief that it shares the key ${g^{xy}}$ with B. The attack does not compromise secrecy of data because B does not learn the session key. Moreover, one could argue that A has correctly authenticated C, and more controversially, that C has correctly authenticated B because B is entitled to choose any key share it likes. Nevertheless, something clearly is amiss about the authentication. A and C have different understanding of who they are communicating with, which violates a correspondence property that an authenticated key exchange intuitively should have.

The above attack was identified by Diffie et al.~\cite{diffie1992authentication} and it has been given many names including \textit{unknown-key-share}~\cite{blake1999unknown} and \textit{identity misbinding}~\cite{krawczyk2003sigma}. In different versions of the attacks, the misled party may be the initiator or the responder or both. Diffie et al.\ initially presented the attack to motivate the station-to-station (STS) protocol. In basic STS, the signatures are encrypted with the Diffie-Hellman session key, and the paper also suggests another variant where a message authentication code (MAC) replaces the encryption. The function of the encryption or MAC is to bind the session key to the signatures, which prevents the attacker, who does not know the session key, from replacing the signatures. 

The STS protocol, including both the encryption and MAC variants, is still vulnerable to misbinding attacks if the attacker B manages to register A's or C's public signature key as its own. This vulnerability is well known and caused by failure of the certification authority to verify that the subject possesses the private key. Nevertheless, the dependence on the CA following best practices can and should be avoided. The SIGMA protocol family by Krawczyk~\cite{krawczyk2003sigma} computes the MAC explicitly on the message sender's identifier, rather than its signature. The SIGMA protocols are highly influential because they include the IKEv2 key exchange~\cite{rfc7296} and its predecessors in the IPsec protocol suite. As a consequence, resistance to the misbinding attacks is considered one of the critical requirements for key-exchange protocols designed for the Internet.

\subsection{Device pairing and relay attack}

Secure \textit{device pairing} is a bootstrapping process that establishes a secure channel between two previously unassociated devices. These devices often communicate over a short-range wireless channel such as Bluetooth~\cite{bluetooth2016}, Wi-Fi~\cite{7786995}, or Zigbee~\cite{alliance2012zigbee}. While the goals of device pairing are similar to those of any authenticated key-exchange protocol, there is one major difference: the devices typically have no prior security context, such as knowledge of each other's public keys or certificates and identifiers. They may not even have identifiers or an assigned owner before the pairing establishes those. Additionally, the devices may not be able to rely on the availability of trusted infrastructure due to the ad-hoc and local nature of the short-range wireless communication. 

Typical device pairing protocols perform a Diffie-Hellman (DH) or an Elliptic Curve Diffie-Hellman (ECDH) key exchange over the in-band wireless channel and then use a \textit{human-assisted out-of-band (OOB) channel} to thwart potential impersonation and man-in-the-middle attackers in the in-band channel. Several researchers have studied the security and usability of device pairing protocols in significant detail~\cite{kainda2009usability,saxena2006secure,gajbhiye2016design,hassan2018security}. The existing literature assumes a powerful Dolev-Yao type attacker on the in-band wireless channel and an OOB channel that provides some inherent protection for the confidentiality and/or integrity of the data exchanged over it. 

Bluetooth (see Section \ref{sec:bluetooth}) is one of the most widely deployed and analyzed wireless technologies. Modern Bluetooth devices use the \textit{Simple Secure Pairing} (SSP)~\cite{bluetooth2016} protocols, although some may be backward compatible with the less secure Legacy Pairing methods. Wireless devices have different input and output capabilities, which is why SSP supports multiple different user interactions and is actually a family of key-exchange protocols. In the \textit{numeric-comparison} mode, the user is asked to compare six-digit codes on two device displays while, in the \textit{out-of-band} mode, the user delivers similar verification information securely from one device to another. Either way, the out-of-band communication by the user prevents man-in-the-middle attacks on the ECDH key exchange that takes place over the in-band wireless channel. There is also a \textit{just-works} mode for devices that support neither output nor input of six-digit codes. Obviously, this mode lacks secure authentication.

Research literature on Bluetooth security discusses several attacks that are relevant to pairing protocols in general. It may be possible to spy on the OOB channel or to misrepresent the device capabilities so that the devices negotiate the insecure just-works mode~\cite{haataja2010two}. The attacker can trick remote devices into believing that they are in direct communication by \textit{relaying} unmodified protocol messages between their locations~\cite{levi2004relay}. In the legacy version of Bluetooth where session encryption was not mandatory, relaying of the authentication messages could result in pairing of the wrong devices. In modern protocols, this attack is relevant when the primary goal is the device authentication and not the following communication, for example, when a Bluetooth device is used as a door key or as a location beacon. The Bluetooth just-works mode can lead to accidental or maliciously induced association with a wrong peer device, as noted among others by Suomalainen et al.~\cite{suomalainen2007security}. If the device supports multiple simultaneous key exchanges, there can be confusion between the resulting sessions~\cite{chang2007formal}. The end result in these attacks is akin to identity misbinding because the reality of the created security associations does not correspond to the device's or user's perception.

Poorly designed internal architecture of a Bluetooth endpoint, such as a mobile phone, may also lead to attacks. Naveed et al.~\cite{naveed2014inside} describe how malicious applications on an Android smartphone can hijack connections from attached Bluetooth (medical) devices in order to steal data. The problem arises from the fact that the Android permission and security model allows any application with the Bluetooth permission to communicate with all external Bluetooth-paired devices. A more general lesson that we can draw from the paper is that it is important to pay attention to malicious insiders, such as untrusted apps, residing in the endpoint devices, which may be able to interfere with the communication without fully compromising the device.

The pairing protocols critically depend on user actions, such as comparing or delivering codes. Ellison~\cite{ellison2007ceremony} introduced the concept of security \textit{ceremonies} where the users are participants to the protocol and their actions are specified, modelled and analyzed just like those of the communicating endpoints. Carlos et al.~\cite{carlos2013updated} use Bluetooth as an example for reasoning about basic security properties of a security ceremony.

\subsection{Trusted computing and cuckoo attack}
\label{sec:trustedcomputing}

The published work closest to ours comes from the trusted-comput\-ing community. In trusted computing, a computer or a mobile device incorporates a secure hardware component that is certified by the manufacturer and acts as a trusted entity inside the device. The most common secure hardware component is a \textit{trusted platform module} (TPM)~\cite{tpm2}, which supervises the boot process of the device and either enforces secure boot or measures (as a cumulative hash value) the loaded software. The latter case is also called dynamic root of trust for measurement (DRTM). The latest microprocessors have more advanced trusted execution environments (TEE), such as ARM TrustZone\footnote{\url{https://developer.arm.com/technologies/trustzone}} and Intel SGX\footnote{\url{https://software.intel.com/en-us/sgx}}, which allow trusted software to be isolated and launched after the device has booted. A common feature in these technologies is that, in addition to enforcing some security policies inside the computer, they can \textit{attest} the integrity of the device and its software configuration to an external verifier. This could allow, for example, the user to cryptographically verify the integrity of a cryptocurrency wallet before storing high-value secrets to it. The attestation naturally needs to be cryptographically linked to a secure communication channel~\cite{goldman2006linking} with the verifier. 

Parno et al.~\cite{parno2011bootstrapping} first pointed out the problem that, while users may be able to cryptographically verify that they are communicating with a trusted hardware module and measured software, it is difficult to be certain that they are physically accessing the very device where that module is embedded. In the \textit{cuckoo attack}, the device in the verifier's physical proximity is not actually trusted but tricks the verifier into believing so. The cuckoo device achieves this by forwarding the communication to another device which has the correct configuration and a DRTM for attesting it. 

Fink at al.~\cite{fink2011catching} suggest measuring the round trip times of requests to the trusted device to detect if it is in the proximity of the verifier. Zhang et al.~\cite{zhang2017presence} also investigate the problem of a human user distinguishing genuine secure hardware from adversarial devices. They divide the presence attestation into two phases: first, existence checking, which uses the standard remote attestation protocols, and second, residence checking, which provides assurance that the attesting hardware module is, in fact, in the specific physical device. We will return to the suggested mechanisms for residence checking in Section~\ref{sec:mitigation}. Ding et al.~\cite{ding2018initializing} further argue that presence attestation with DRTM differs significantly from device pairing where both devices are trusted. The current paper sets out to investigate whether this is always the case.

\subsection{Formal modelling}

Formal modelling and model checking are standard methodology in the development and analysis of key-exchange protocols~\cite{blanchet2001efficient,armando2005avispa,dill1996mur}. Various protocol flaws have been found with these methods but, perhaps more significantly, formal models are a way to lift the security-protocol design to a higher abstraction level than message formats and state machines, and to define precisely the security properties that the protocol is expected to have. 

The model checkers for security protocols are special compared to other formal modelling tools in that, in addition to taking the system design as input, they typically have a built-in model of the Dolev-Yao type powerful attacker, which the researcher does not need to explicitly define. Instead, the researcher has to specify the desired security properties. The model checker then determines whether the attacker is able to play a game against the honest parties and trick them into violating these properties. There is, however one type of attack that the researchers need to explicitly consider: corrupt insiders. The corruption of an insider is often modelled as a previously honest party handing out its secrets and capabilities to the attacker, after which it is subsumed into the attacker.

Jia and Hsu~\cite{jiaformal} develop a formal model of the Bluetooth SSP for the Murphi model checker~\cite{dill1996mur}. They discuss two potential vulnerabilities in the numeric-comparison authentication mode. First, an \textit{impersonator} device can pretend to be a good one and trick the user into pairing an honest initiator device with it. The example given in the paper is one where the entertainment system in a rental car has been replaced with one that is under the adversary's control. Once the unsuspecting user has paired her phone with it, the system can steal confidential data. Second, a \textit{proxy MitM} device can forward the unmodified connection to another device (similar to \cite{levi2004relay}). While these threats might be considered obvious and unavoidable, the formal analysis focuses our attention to them and enables systematic consideration of the threats.

The most interesting idea of Jia and Hsu for us is the notion of \textit{intention preservation}. It means that the initiating device is paired with the device with which the user originally intended to pair it, even if the non-initiating device belongs to an intruder. They show that Bluetooth pairing with numeric comparison has this property. We develop further the idea of modelling user intention, which we state as a correspondence assertion. Because of subtly different security definitions, we end with a different result regarding Bluetooth pairing. 

%% file: inpairing.tex

\section{Misbinding in device pairing}
\label{sec:inpairing}

\begin{figure}[tb]
\centering
\includegraphics[scale=0.5]{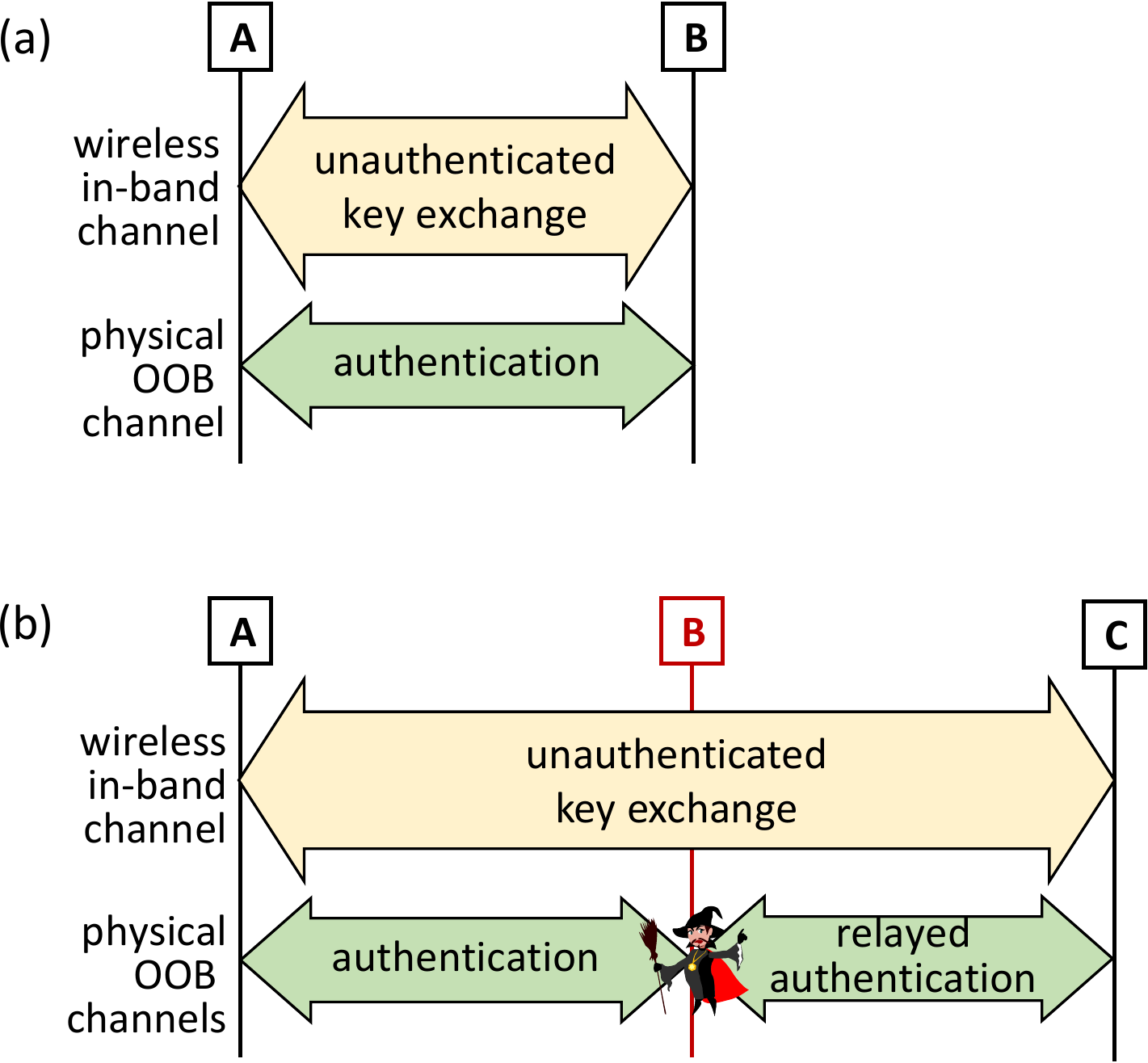}
\caption{Identity misbinding against device pairing}
\Description{Identity misbinding attack against device pairing, a high-level view}
\label{fig:misbinding-pairing}
\end{figure}

We will now look at identity misbinding attacks against wireless device pairing where user authenticates the key exchange between two physical devices. Figure~\ref{fig:misbinding-pairing}(a) shows a common structure for many such pairing protocols. The unauthenticated key exchange takes place over an insecure in-band channel, and the user with physical access to the devices authenticates it over a secure out-of-band channel. The two phases may not always be distinguishable by time, but they are distinguishable by the channel. 

The authentication in user-assisted pairing protocols is typically based on physical access to the device. That is, the user must see or touch the devices directly. The devices could have serial numbers, public keys, or other unique identifiers, but it is the physical access that defines which devices need to be paired. 

We consider a scenario where one of the devices selected by the user for the pairing is compromised. (Recall that identity misbinding is an insider attack where one of the intended communication endpoints is corrupt.) The device has to be compromised at least to the extent that the user can control the device's inputs and outputs on the OOB channel. In Figure~\ref{fig:misbinding-pairing}(b), the user wants to pair devices A and B. However, device B is malicious and relays the authentication messages to another device C. Devices A and C end up paired, which does not correspond to the user's intention. Device C does not need to collude with B and may be entirely honest, except that the attacker can put it into the pairing mode and interact with it.  

Let's try to understand why this attack is not easy to prevent. If we take guide from other authenticated key-exchange protocols, such as STS and SIGMA, we might try to prevent the attack by binding the endpoint identifiers A and B cryptographically to the key exchange and the created session. This will ensure that the endpoints of the created session agree on the identities. Sadly, that does not help in device pairing. The attack by B will cause A and C to be paired, but if the user is not aware of the identifiers communicated in band, the user still thinks A is paired with B. As the next step towards a solution, we would need to check that the device identifiers A and B correspond to the user's expectations. For example, if device A shows the peer identifier to the user, the user sees that it is C and not B as intended. However, the typical user in device pairing does not have any expectations about the device identifiers: the user just sees two physical devices and wants them to be paired.  

Many pairing protocols are like this: the user's physical access to the device defines its identity. Since the physical device identity cannot be communicated in bits and bytes, it cannot be included into the messages sent over the in-band or out-of-band channel, and it cannot be used as input to a cryptographic function. Cryptographic protocol vulnerabilities of the early days could often be fixed by adding a missing identifier to the right message, but that is not the case with device pairing where the endpoints either have no identifiers or, if identifiers exist, user intentions are not expressed in terms of them.

So far, our discussion of misbinding may appear as rehashing of the relay attack in the context of device pairing. This perception is partly true, but the misbinding attack is actually far easier to implement. As hinted in Figure~\ref{fig:misbinding-pairing}(b), if all three devices are within the wireless range from each other, B does not actually need to relay the wireless in-band traffic. It can let A and C communicate directly over the wireless channel and focus on relaying the authentication messages between the two OOB channels. B can then pull out after the authentication is complete, which leaves A and C communicating directly. 

Comparing with the cuckoo attack against trusted computing hardware, there are also similarities. The problem there was the lack of secure binding between the physical device and the long-term public key of the DRTM inside it. Our problem is the lack of secure binding between the physical devices and the ephemeral session key. The similarity extends to the lack of definite solutions by the means of traditional security protocol design. However, there are ways of mitigating the threats, as we will see in Section~\ref{sec:mitigation}. 

Next, we will look at some examples of the attack in actual pairing protocols. That will help us assess the impact of the vulnerability in a more concrete way. 

%% file: bluetooth.tex

\subsection{Bluetooth case study}
\label{sec:bluetooth}
\begin{figure}[tb]
\centering
 \includegraphics[scale=0.5]{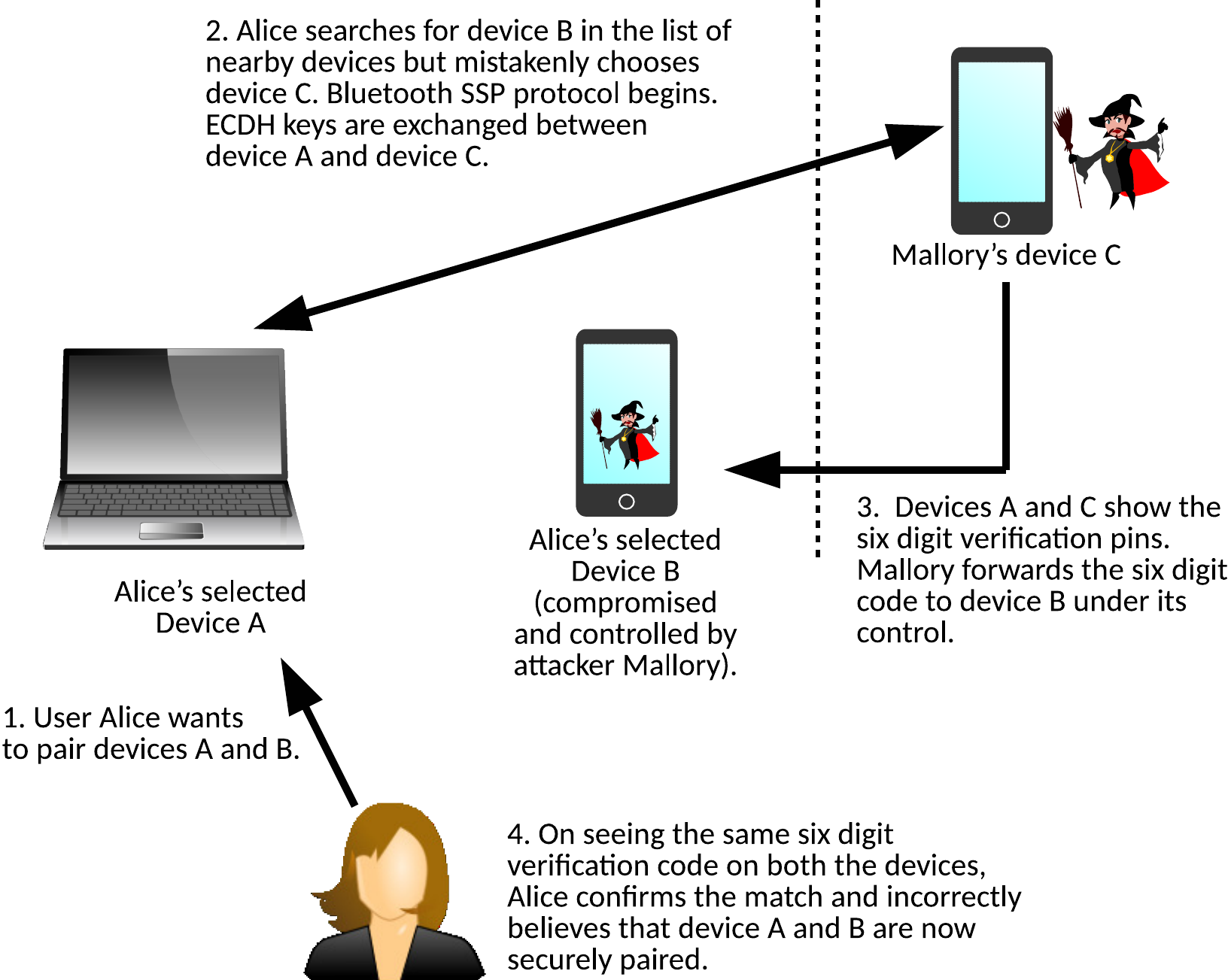}
\caption{Misbinding attack against Bluetooth SSP numeric comparison}
\Description{Misbinding attack against Bluetooth Secure Simple Pairing with numeric comparison}
\label{fig:bluetooth-misbinding}
\end{figure}

We use the widely-studied Bluetooth SSP as a case study of misbinding in pairing protocols. The attack is shown in Figure~\ref{fig:bluetooth-misbinding}. The human user Alice is unaware of the fact that the device B, to which she is trying to pair her phone A, is compromised and under the control of an attacker Mallory. The attacker also has a third device C, which she keeps hidden from the user. The attacker's goal in the misbinding attack is to pair Alice's device A with the third device C while Alice believes A is paired with B. For a successful misbinding attack, A and C must be within Bluetooth radio range from each other. For example, Mallory and device C could be in the next room from where Alice performs the pairing process.

From the user's and the attacker's points of view, the following steps occur in the misbinding attack of Figure~\ref{fig:bluetooth-misbinding}: 
\begin{enumerate}
\item Alice wants to pair devices A and B with the goal of establishing a secure association between them. Alice is unaware of the fact that device B is compromised and that a third device C, accessible by the attacker Mallory, is within radio range.
\item Alice starts a search for new Bluetooth devices on device A. She makes device B discoverable, if it is not yet so. Mallory makes device C discoverable. Device A then presents Alice with a list of the names of discoverable devices in its vicinity. Alice chooses the one she thinks is B. At this point, Mallory needs to arrange things so that Alice mistakenly chooses C from the list. To achieve this, Mallory should ensure that the compromised device B remains non-discoverable, even though Alice thinks otherwise, and ensure that the name of device C matches the name that Alice expects to see for device B. (We will discuss the naming in more detail below.)
\item During the pairing, devices A and C show six-digit codes and expect the user to compare them. Mallory reads the six-digit code from the screen of device C and forwards it to the compromised device B, which displays it to Alice.
\item Seeing the same six-digit verification code on the screens of devices A and B, Alice confirms the pairing on both devices. The action on the compromised device B has no real effect; instead, Mallory confirms the pairing on device C. This allows the pairing of A and C to complete. Alice now believes A and B have successfully paired when, in fact, device A is paired with C. 
\end{enumerate}

To understand why the Bluetooth SSP protocol does not prevent the attack above, we need to look at the protocol in more detail. The hardest practical obstacle for the attacker is, in fact, not the actual SSP protocol but the device naming and selection that takes place before the actual pairing. Bluetooth core specification~\cite{bluetooth2016} defines Inquiry and Paging procedures for discovering nearby devices and subsequently connecting to one of them. The user typically selects the name of the non-initiating device from a list of nearby devices on the initiating device. The device names are strings that aid the user in identifying the correct peer device. Each device has a default name that often indicates its make and model, for example ``TomTom Go 510''. Depending on the device, the name may be user configurable. In the attack, Mallory needs to trick Alice into choosing device C from the list by its name. Thus, Mallory should rename C to have the same name as B. 

The rare tricky case for Mallory is if she wants to use a device C that does not have a configurable name, or if Mallory does not have the permission to change the device name. In that case, Mallory may be able to choose a device C that has the same make and model as device B and thus the same default name. If Mallory absolutely needs to use a device C with a Bluetooth name that is not configurable and does not match device B, there is still a way forward. The Inquiry and Paging procedure is not authenticated, and the attacker can manipulate the device names on the in-band wireless channel. While that requires more skill than changing the name of device C on its user interface, message modification on a wireless channel is within the expected capabilities of a Dolev-Yao attacker.

\begin{figure}[tb]
\centering
 \includegraphics[scale=0.5]{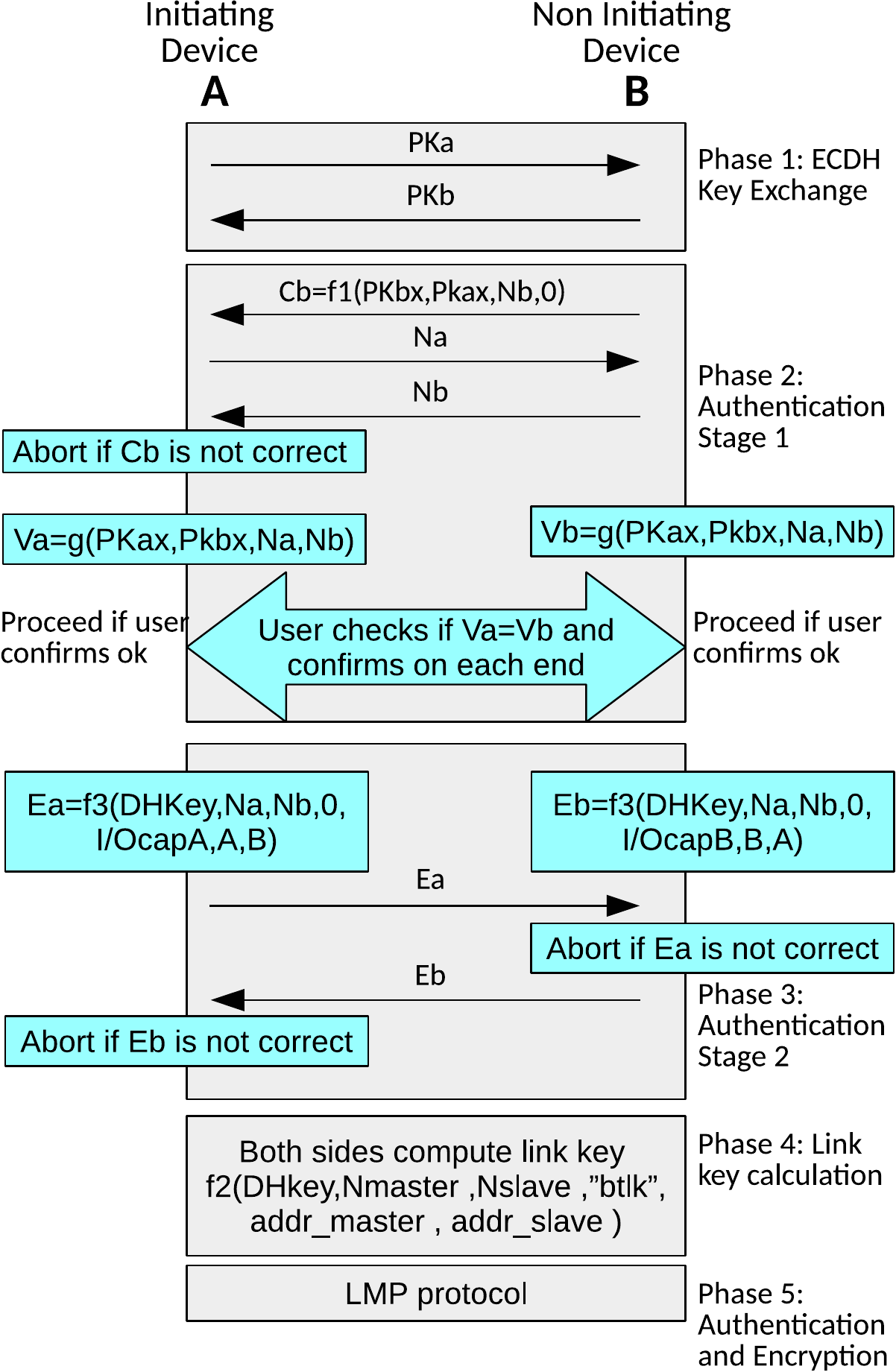}
\caption{Bluetooth Secure Simple Pairing with numeric comparison~\cite{bluetooth2016}}
\Description{Message sequence chart of the Bluetooth Secure Simple Pairing with numeric comparison}
\label{fig:bluetooth}
\end{figure}

Once Alice has been fooled into choosing the wrong device, the SSP security protocol starts between devices A and C. We will review the protocol to be certain that it does not present obstacles to the attack. The numeric-comparison mode of SSP, shown in Figure~\ref{fig:bluetooth}, has several phases that must be completed before an initiating device A and a non-initiating device B are paired securely. In phase 1, the devices perform an ECDH key exchange. In phase 2, the non-initiating device commits to a random nonce Nb, which it reveals after the initiating device has sent its own nonce Na. Device A checks the commitment to ensure that the nonces have been fairly chosen. The user-assisted authentication then takes place. Each of the devices displays to the human user a six-digit verification code, which it computes from the ECDH key shares and nonces. If the codes match, the user confirms successful pairing on both devices, which allows them to continue. In phase 3, the devices confirm cryptographically the derived ECDH secret and their input and output capabilities, which were used to select the authentication mode in the beginning. In phase 4, the devices derive a link key, i.e.~a shared session key. Finally, in phase 5, they use the link key for encryption in the Link Manager Protocol. 

The critical thing to observe about the SSP protocol is that it does not even try to verify the device names (or other device properties like make, model and serial number). This is understandable because Bluetooth device names do not uniquely identify a device. The protocol does bind the link key to the link-layer addresses of the two devices, but during the pairing each device will accept any peer address.

Note that only the software in device B needs to be compromised for the misbinding attack, while devices A and C can be entirely normal. The only access the attacker needs on device C is to make it discoverable, to change its name if necessary, and to confirm the code comparison. Moreover, the attack requires device B to be compromised only to the extent that the attacker can control its user interface. We implemented the attacker in device B as a full-screen app that receives the six-digit code over the 4G data connection and emulates the pairing process without doing actually anything. Thus, the vulnerability occurs relatively often in practice, even though we do not know of actual attack implementations outside our laboratory.

The above attack against Bluetooth pairing will work for any version of SSP or Legacy Pairing. Indeed, we believe it will work for all device-pairing protocols where the device identity is determined by physical access to the device alone.

%% file: eapnoob2.tex

\subsection{IoT device bootstrapping case study}
\label{sec:eap-noob}

We will now look at a protocol for security-bootstrapping and registration of  Internet-of-Things (IoT) devices to an online server. Although the protocol differs considerably from device pairing, they are similar in the sense that the identity of the correct device is defined by physical access to it. This makes the protocol vulnerable to identity misbinding attacks. 

Extensible Authentication Protocol (EAP)~~\cite{rfc3748} is an authentication framework used, for example, in enterprise wireless networks. It normally assumes that the wireless devices are pre-registered at a back-end authentication server. This means that the deployment of new wireless devices is a multi-step process that includes device registration and credential provisioning. 

Nimble out-of-band authentication for EAP (EAP-NOOB)~\cite{aura-eap-noob-05} is an authentication method for EAP that also supports user-assisted bootstrapping and registration of new devices. It is intended for off-the-shelf IoT devices that initially have no known identifiers, no credentials, and no knowledge of their intended owner and network. EAP-NOOB registers the new devices to the authentication server and associates them with the user's account on the server. The device, called \textit{peer}, first performs an ECDH key exchange with the server. The authentication takes place when the user delivers a single out-of-band (OOB) message from the peer device to the server, or in case of peer devices with only input capability such as cameras, from the server to the peer device. Information delivered in the OOB message enables mutual authentication of the peer and server, and it authorizes, on one hand, the server and user to take control of the device and, on the other, the device to be registered to the server and user account. The protocol does not limit the ways in which the OOB message is transferred; the implemented ways include a QR code, an NFC message, and an audio clip. After the OOB message has been delivered, the device registration completes in-band between the peer and the server.

\begin{figure*}[!t]
\centering
    \includegraphics[scale=0.5]{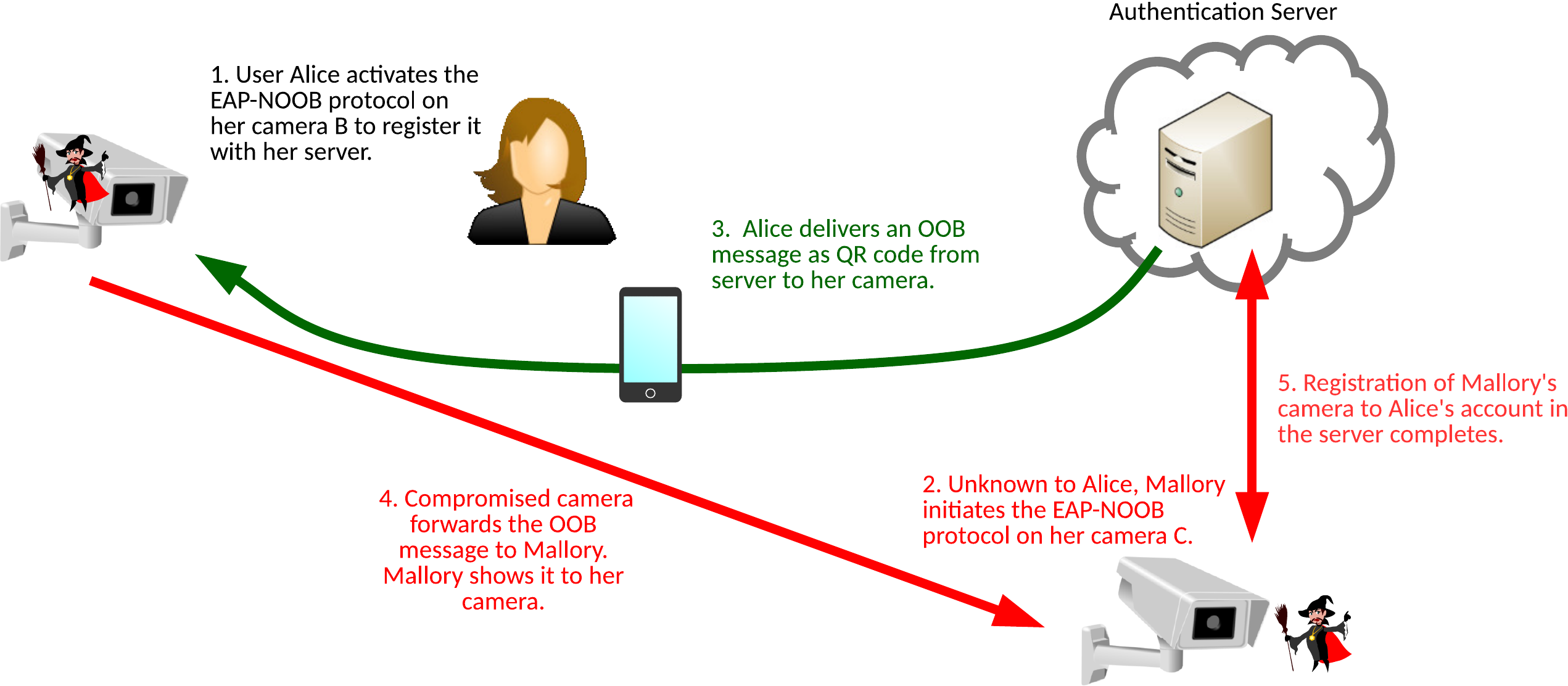}
    \caption{Misbinding attack against EAP-NOOB}
    \Description{Misbinding attack against new-device registration with the EAP-NOOB protocol}
    \label{fig:noob-misbinding}
\end{figure*}

The misbinding attack (shown in Figure~\ref{fig:noob-misbinding}) arises when the peer device B is compromised. It can trick the user into registering a different peer device C to the user's account in the server. From the user's and the attacker's points of view, the following steps occur in the attack:
\begin{enumerate}
\item Alice initiates the registration of her web camera B to the wireless network and authentication server. Unknown to Alice, camera B is compromised and under the control of Mallory. Camera B pretends to start the EAP-NOOB protocol with the server. 
\item At the same time, Mallory initiates the registration of another web camera C to the same network and authentication server. Camera C starts the EAP-NOOB protocol with the server. 
\item Alice logs into her user account on the server with her mobile phone and searches for new cameras available for registration. Finding a new camera that matches the model of camera B, she retrieves a QR code encoding the OOB message and shows it to camera B. 
\item The compromised camera B scans the QR code and secretly passes a picture of it to Mallory. Mallory shows the QR code to her camera C. This authorizes the registration of camera C to Alice's account on the server. 
\item The web camera C and the server now complete the registration of camera C to the authentication server, associating it with Alice's account, and establish credentials for future authentication and wireless network access. Alice mistakenly believes that the new camera associated with her account is B, when in fact, it is C.
\end{enumerate}

In order to trick Alice into selecting the wrong camera from the server, the attacker needs to match its make and model or other metadata that the user is likely to search for. The attacker can achieve this by using another camera of the same type. In that case, the attacker does not need to modify camera C in any way. She must be able to start the EAP-NOOB registration process on device C, but this is not difficult: the EAP-NOOB specification suggests that the protocol could be activated by powering up the device for the first time or after a hard reset. An alternative for the attacker is to have an entirely fake device C that is under her control and spoofs the device metadata copied from B. It could even copy the serial number of device B. 

Unlike in device pairing, Mallory's device C does not have to be in close proximity to Alice's device B. Mallory can run the EAP-NOOB protocol on her device C from anywhere in the coverage area of the wireless networks served by the same authentication server. She only needs the capability of sending or receiving the OOB message to or from the compromised device B.

Device bootstrapping and registration with EAP-NOOB is designed to be efficient for deploying large numbers of devices. Thus, the person installing the devices might not be the eventual user, and the failure of device B to associate with the server might go unnoticed for some time. In comparison, device pairing with Bluetooth is often followed by another user action such as transfer of media, which may lead to the user detecting the failure of device B to pair.

%% file: proverif.tex

\section{Formal analysis of misbinding}
\label{sec:proverif}

We modelled the case-study protocols and their security requirements with ProVerif~\cite{blanchet2001efficient,blanchet2018proverif}. First, we wanted to enhance previous models of device pairing and especially Bluetooth SSP to capture the misbinding attack. It was not clear to us why the existing models missed the attack when so many other, even more subtle issues have been detected. We also wondered if the attack and the security goals it violates can be reduced to previously known ones. As a result, we learned that the formal models can be made more complete so that they discover the misbinding attack, and that the violated security properties are different from what has previously been analyzed. Another goal of our modelling work was to understand how pairing protocols differ from each other in relation to the misbinding vulnerability, and whether registering a physical device to an online service is fundamentally different from pairing two physical devices. We found that misbinding occurs in a wide range of protocols where endpoints are defined by physical access. We also found that the attacks can be classified into a small number of variants, and not all protocols are vulnerable to all of them.

\subsection{Modelling device pairing}
\label{sec:proverif-bluetooth}

We will mainly discuss Bluetooth SSP with numeric comparison because of its familiarity to many readers. However, we also modelled the SSP OOB mode and Wi-Fi Direct~\cite{wifi-direct} with similar results. 

In addition to the protocol messages and the device state machines, we model the \textit{security ceremony that includes user intentions, choices and actions}. We follow the example of Carlos~\cite{carlos2014towards} and model the user as a separate process in ProVerif. However, while Carlos considers pairing between two devices belonging to different users, we consider pairing where a single user has physical access to both intended endpoints. Thus, our model consists of three kinds of processes: user, initiating device A, and non-initiating device B.

The challenging part of the model was capturing the user intention, i.e.~decision to pair specific two devices, when the \textit{devices are identified by physical access and do not have names or other identifiers}. 
In the end, the solution is fairly simple and intuitive: the users and devices have identifiers in the model (see below), but the identifiers can never be communicated over a channel or used as input to a cryptographic function. Instead, they are used for marking local events and for checking correspondence properties between the events, such as whether the user intended the devices to be paired. This inability to communicate the identifiers goes a long way towards explaining why the traditional solutions of adding explicitly or implicitly communicated identifiers are not applicable to device pairing. 

Similar to Chang et al.~\cite{chang2007formal}, we use private channels in ProVerif to model the \textit{physical access} by the user to the devices. These channels protect both secrecy and integrity of the communication. In the case of Bluetooth, the private channels are used both for reading the numeric codes and, if the values match, for confirming the match to the devices. To initiate pairing, user needs to have access to two private channels, \texttt{\small PhysicalChannelA} to an initiator device and \texttt{\small PhysicalChannelB} to a non-initiator device. 
We use these physical channels as the device identifiers, which is both practical and semantically correct. For the users, on the other hand, we simply create new identifiers.

Compromised endpoints are commonly modelled by leaking their secrets, such as private keys, to a public channel. Consequently, the built-in attacker model of the model-checking tool can emulate any honest or malicious behavior by that endpoint. In the Bluetooth model, however, the devices do not have any master secrets. Instead, we \textit{model the compromise of a device by leaking its private channel to the network}. This allows the attacker to take control of that channel.

In addition to modelling the compromise of devices, we also model the compromise of a user. This is done to conceptually distinguish between a tampered device and a malicious user having physical access to an intact device. There is no real difference between the two in the Bluetooth case.

The user model is shown below. The user \begin{inparaenum}[(i)] \item selects two devices and logs her decision to pair them as an event, \item compares the six-digit verification codes displayed by the devices, and \item confirms a match to the devices\end{inparaenum}. The user may be compromised any time, yielding control of the physical access channels to the attacker.

\begin{lstlisting}[
    breaklines=true,
    basicstyle=\footnotesize
]
let UserProcess(User:User_t, PhysicalChannelA:channel, PhysicalChannelB:channel) =
(
  event HasAccess(User, PhysicalChannelA);
  event HasAccess(User, PhysicalChannelB);
  (* Decide to pair A and B with A as initiator *)
  event IntendToPair(User, PhysicalChannelA, PhysicalChannelB);
  (* Receive Va and Vb *)
  in(PhysicalChannelA, (=CodeTag, Va:Hash_t));
  in(PhysicalChannelB, (=CodeTag, Vb:Hash_t));
  (* Numeric comparison *)
  if Va = Vb then
    (* Confirm to A *)
    out(PhysicalChannelA, (OkTag, Va)); 
    (* Confirm to B *)
    out(PhysicalChannelB, (OkTag, Vb))  
) | (
  event CompromiseUser(User);
  out(c, PhysicalChannelA);
  out(c, PhysicalChannelB)
).
\end{lstlisting}

Intuitively, misbinding is a violation of the following security property: \textit{two devices are paired only if their user intended them to be}. When formalizing the absence of misbinding as a correspondence property in ProVerif, we need to be more precise: \textit{If two devices complete the pairing with the same link key and a user has physical control of at least one of them, then either the user previously intended the two devices to be paired, the user is compromised, or both devices are compromised.} In ProVerif, this correspondence property can be defined as follows:

\begin{lstlisting}[
    breaklines=true,
    basicstyle=\footnotesize
]
query PhysicalChannelA:channel, PhysicalChannelB:channel, K:Key_t, User:User_t;
  (event(HasAccess(User, PhysicalChannelA)) && (*or B*)
   event(InitiatorComplete(PhysicalChannelA, K)) &&
   event(NoninitiatorComplete(PhysicalChannelB, K))) 
  ==>
  (event(IntendToPair(User, PhysicalChannelA, PhysicalChannelB)) ||
   event(CompromiseUser(UserId)) ||
   (event(CompromiseDevice(PhysicalChannelA)) && 
    event(CompromiseDevice(PhysicalChannelB)))).
\end{lstlisting}

As expected, ProVerif returned \textit{false} for the query and produced a counterexample, i.e.~an execution trace that violates the security property. There are two versions of the query, one with \texttt{\small Physical\-ChannelA} and another with \texttt{\small PhysicalChannelB} on the second line. The queries can be refined to exclude already analysed attacks or to focus on specific cases.

\begin{figure}[!h]
\includegraphics[scale=0.42]{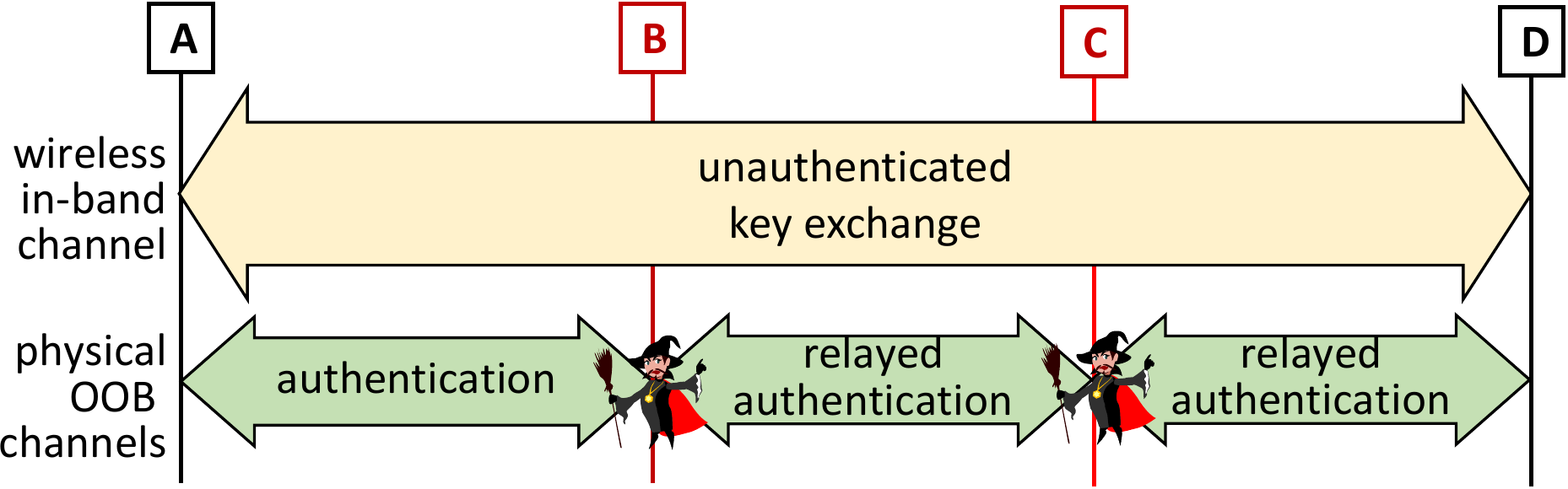}
\caption{Double misbinding}
\Description{Double misbinding attack where two uncompromised devices paired without the two honest users' knowledge}
\label{fig:misbinding-double}
\end{figure}

\begin{figure*}[tb]
\includegraphics[scale=0.41]{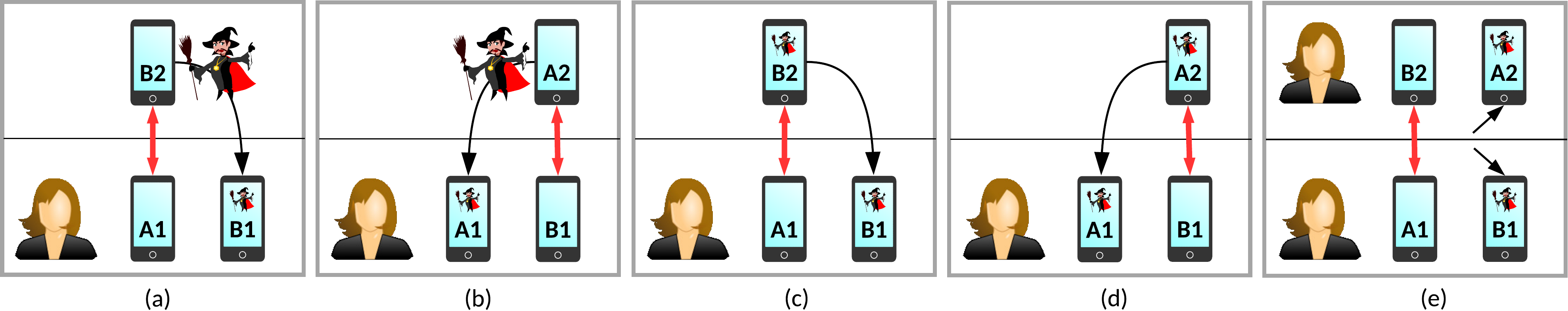}
\caption{Five variants of misbinding found by ProVerif}
\Description{Five variants of the misbinding attack found with ProVerif}
\label{fig:five}
\end{figure*}

Investigating further, we found five different types of misbinding attacks with ProVerif. One of them is the basic misbinding attack described in Section~\ref{sec:inpairing} and shown in Figure~\ref{fig:misbinding-pairing}(b). In that attack, the compromised device is the non-initiator B, and there is a compromised user with physical access to the third device C. Other attacks arise as variations of the first one: On one hand, the compromised device can be the initiator A or the non-initiator B. On the other hand, device C may be a compromised one or an uncompromised device accessed by a compromised user. These choices make four different variants of the misbinding attack.

It came as a surprise to us that there is a fifth type of misbinding attack, which we call \textit{double misbinding}. In this attack, shown in Figure~\ref{fig:misbinding-double}, there are two honest users. Each one of them is trying to pair two devices, one of which is compromised. The compromised devices collude so that, as the end result, the two uncompromised devices are paired.

Double misbinding is easiest to understand in the out-of-band mode of Bluetooth SSP, where the user transfers some information OOB from one device to another. In that case, one compromised device receives the OOB message from the first honest user and forwards it secretly to the second compromised device, which outputs it to the second honest user. The attack is also possible in SSP with numeric comparison because all the values needed for computing the verification codes Va and Vb are transmitted on the wireless link (see Figure~\ref{fig:bluetooth}). The attacker can sniff these values, compute Va and Vb, and show them on the displays of the two compromised devices (devices B and C in Figure~\ref{fig:misbinding-double}). 

The five variants of misbinding are summarized in Figure~\ref{fig:five}. Each sub-figure shows two rooms. The honest user tries to pair two devices, initiator A1 and non-initiator B1, in her room but one of them ends up being paired (indicated by the thick red arrow) with a device in the room above. The sub-figures show the locations of the honest users, compromised users, and compromised devices. The black one-directional arrow is specific to Bluetooth SSP with numeric comparison. It shows how the attacker forwards the six-digit code from one device to another or, in the double-misbinding case of Figure~\ref{fig:five}(e), sniffs its inputs from the wireless communication. 

Afterwards, we systematically enumerated the different combinations of initiator and non-initiator devices, compromised and uncompromised users and devices, and user physical access in a setting of maximum two users and four devices. This analysis confirmed that, after removing impossible and equivalent cases, the five attack variants remain. Increasing the number of users and devices does not seem to give raise to any new types of attacks because there is maximum that can be involved in a single pairing.

\subsection{Modelling device bootstrapping}

Although the ProVerif models of EAP-NOOB and Bluetooth differ greatly, the parts relevant to detecting misbinding are similar. The main difference is that, in EAP-NOOB, only the peer device is identified by the physical access channel. The EAP-NOOB server has a strong cryptographically verifiable identity (HTTPS URL and web certificates), and we assume that the server cannot be compromised. The query for the absence of misbinding attacks is as follows:
\begin{lstlisting}[
    breaklines=true,
    basicstyle=\footnotesize
]
query OobChannelS:channel, OobChannelP:channel, K:Key_t, User:User_t;
  (event(HttpsAccess(User, OobChannelS)) && 
   event(ServerComplete(OobChannelS, K)) &&
   event(PeerComplete(OobChannelP, K))) 
  ==>
  (event(CompromiseUser(User)) ||
   event(IntendToPair(User, OobChannelS, OobChannelP)) ||
   (event(CompromisePeer(OobChannelP)) && 
    event(CompromiseServer(OobChannelS)))).
\end{lstlisting}

Again, ProVerif finds a counterexample to this query. Because only the peer side can be compromised, there are only two possible variants of misbinding. One is the attack of Figure~\ref{fig:bluetooth-misbinding}(b) with server as A, compromised peer device as B, and an uncompromised peer device as C. In the other attack variant, both peer devices are compromised and there is no need for a user to operate device C. These variants correspond to Figure~\ref{fig:five}(a) and \ref{fig:five}(c) if we interpret A1 as the authentication server, B1 as Alice's wireless device, and B2 as the attacker's device.

%% file: mitigation2.tex

\section{Mitigation}
\label{sec:mitigation}

\subsection{Authentication solutions} 

As explained in Section~\ref{sec:identitymisbinding}, the STS and SIGMA protocols and their variants~\cite{diffie1992authentication,krawczyk2003sigma,blake1999unknown} tackle misbinding by binding endpoint identities cryptographically to the created session. These solutions are suitable for situations where the devices have certificates, public keys for authentication, and unique names. This is typically not the case in device pairing. Moreover, as we explained in Section~\ref{sec:inpairing}, the endpoints in device pairing have no a-priory knowledge of each other's identifiers, and neither does the typical user who is assisting the key exchange. 

The common way to communicate the device identifier, such as model and serial number, to the user is printing them on an identification plate attached to the device. Together with a certificate issued by the manufacturer, this information can be used for authenticating the device. Another possibility is to print a fingerprint of the device's public-key onto the device, e.g.~as a hexadecimal value. If a metal plate, sticker or printing on the device is not considered tamper-proof enough, the identifiers could be etched to the device enclosure. While such physical indicators can ultimately be counterfeited, the burden on the attacker is increased significantly. The disadvantage of these solutions is that the user needs to compare the authenticated device identifiers with the serial-number plates or key fingerprints, which complicates the pairing process.

\subsection{Presence checking}
 
As noted in Section~\ref{sec:trustedcomputing}, trusted-computing research has not put much faith in the printed serial numbers or public-key fingerprints. Instead, the researchers have tried to find more secure ways of checking the presence of a DRTM inside a physical device. We can generalize these approaches from DRTM to any device with a trusted computing base (TCB) that is surrounded by potentially compromised layers of software. The techniques for DRTM presence checking could be applied to checking the physical presence of the pairing endpoint for a given device, which could prevent the misbinding attacks.

The round-trip time measurement suggested by Fink~\cite{fink2011catching} depends on the latency caused by the cuckoo in the communication chain. In our attacks against device pairing, the in-band communication takes place directly with the third device, and timing measurement is unlikely to be able distinguish between two devices within the Bluetooth radio range. This issue of \textit{distance bounding} has been widely studied in relation to RFIDs and wireless keys~\cite{hancke2005rfid}\cite{rasmussen2010realization}. 

Ding et al.~\cite{ding2018initializing} provide a summary of several other solutions. One is a hardware-based secure channel, i.e.~a \textit{trusted path}, that allows the user to communicate directly with the DRTM or TCB inside the device. This could, for example, be an LED indicator light or a special-purpose USB port. The need for such a feature in smart devices is well known, but the idea has never been widely adopted by device manufacturers. The great variety of manufacturers and form factors in smart devices would also make it difficult for the user to know which feature can be truly trusted. Another solution is to enclose the devices into a Faraday cage to prevent them from communicating with external entities during the key-presence checking. This approach was previously suggested for bootstrapping sensor nodes wirelessly~\cite{kuo2007message}. Zhang et al.~\cite{zhang2017presence} propose several presence checking methods based on analog channels, which do not provide strong security guarantees but make the attacks impractical. One method is based on comparing the GPS location measurements by the two endpoints, and another on comparing images captured by co-located devices of their immediate environment. They also propose measuring the timing of a screen-to-camera video channel, which would be difficult to forward to a remote device without causing a detectable delay.

\subsection{Asset tracking}

We believe the practical approach to detecting misrepresented device identities might be \textit{asset tracking}, i.e.~bookkeeping of the physical assets that belongs to an organization or an individual. This requires each device to have a unique identifier, which is registered into a database when the user purchases a device. In the simplest case, the database is accessed only by human users, in which case any existing asset tracking system or database can be used. 

When the organization knows the models and serial numbers of its devices and the purpose assigned to each one, the information can be used for cross-checking during device pairing. For example, if there is only one new display device allocated for Alice, Alice can compare the device information from the database with the identifier authenticated in the device pairing process when she deploys the device. 

For this to work, each device needs to know its own identifier and learn the peer identifier during the key exchange. The identifiers should be bound to the cryptographic key exchange in such a way that agreement on session key cannot be reached without also agreeing on the identifiers. Each device should show the identifier of its peer to the user, e.g.~when initiating the pairing protocol or when confirming the numeric comparison. In Bluetooth SSP protocol, this would require changes to the input of the verification codes, while EAP-NOOB already has a built-in authenticated message field (\texttt{PeerInfo}) for communicating such auxiliary peer information. Of course, the software of an uncompromised device should not allow the users to modify the device identifier. As the result of these measures, device A in the scenario of Figure~\ref{fig:misbinding-pairing}(b) would show the identifier of the unknown device C to the user and the attacker cannot replace it with the expected identifier of device B.

Manufacturer-issued device certificates~\cite{5367679,digicert} can further help the process by providing secure information about the types and models of the devices. This will reduce the reliance on the asset database because all other information except correctness of the device identifier can be communicated in the certificate. 

For the average consumer, it is difficult to keep track of purchased devices over any longer span of time. However, this obstacle may be disappearing as smart devices are increasingly cloud connected and their ownership is therefore often registered by the manufacturer or some other cloud service. The same online service can replace the corporate asset-tracking system for an individual user. Furthermore, there are proposals for logging Internet-of-Things devices to a blockchain~\cite{nuss2018towards,kravitz2017securing}, which could also be used for asset tracking.

Above, we have mostly discussed device pairing and Bluetooth, but the same solutions also work for EAP-NOOB and device registration to the cloud. The main difference is that only one endpoint of the key exchange is a physical asset that needs to be tracked. The fact that the authentication server is online and provided to the user as a service means that it could help with ownership tracking or connect directly to the manufacturers on the user's behalf.

\subsection{On Bluetooth SSP and double misbinding}

As noted in Section~\ref{sec:proverif-bluetooth}, SSP with numeric comparison is vulnerable to double misbinding because all the inputs for computing the verification codes Va and Vb are transmitted on the wireless link and can be sniffed. If Va and Vb were computed as function of the ECDH shared secret, the two compromised devices could not show the value on their displays. This would prevent double misbinding, although not the simpler misbinding attacks. Similar protocols in the future might consider taking advantage of the secrecy to limit the space that the attacker has for maneuvering. A possible disadvantage is that the devices would have to compute the ECDH shared secret before displaying the verification codes, which could impact the user experience on devices with slow processors. The current SSP protocol also has a clean design where the six-digit verification codes are not at all expected to be secret.

%% file: discussion.tex

\section{Discussion}
\label{sec:discussion}
 
It remains to be discussed how serious the misbinding vulnerability is and whether we should really be worried about it. We do not want to be alarmist but instead try to provide balanced arguments for thinking about the issue. 

First, the vulnerability is not specific to Bluetooth SSP and EAP-NOOB, the examples discussed in this paper. To back this claim, let us briefly consider another prominent bootstrapping protocol. Device Provisioning Protocol (DPP)~\cite{DPP} is a bootstrapping mechanism recently standardized by the Wi-Fi Alliance for configuring Wi-Fi network information on devices with limited user interfaces. DPP relies on a \emph{configurator}, e.g. a smartphone application, for bootstrapping all other devices, called \emph{enrollees}, in the network. Every enrollee must have an asymmetric key pair, which is communicated to the configurator over an out-of-band (OOB) channel together with communication metadata such as the radio channel and band. The misbinding attack against DPP is almost trivial: when the user is configuring a compromised device B, the attacker simply replaces the public key and communication metadata output from B with those of another device C. In one variant of DPP, public key is printed as a QR code, and in that case, the device compromise is equal to replacing this piece of paper. 

Any pairing or bootstrapping protocol that relies solely on the user's physical identification of the endpoints will be equally vulnerable regardless of the protocol design. In fact, even strong authentication of the endpoint identifiers does not prevent misbinding unless each endpoint knows what the other's identifier should be. 

One factor that increases the risk of misbinding attacks against out-of-band authentication is that they are easy to implement. The compromised device only needs to forward authentication messages on the user-interface level. Compare this to relaying communication at the radio or logical link layer, or to forwarding application messages. This makes misbinding an attractive attack for technically less competent attackers. In our attack implementations against Bluetooth or EAP-NOOB, the compromise of device B meant simply installing a malicious app that emulated the pairing user interface at the attacker's command. 

Misbinding depends on the user trying to pair with or register a device B that is compromised. Thus, there must be a (partially) corrupt insider involved. The user is misled because the user makes a bad decision and trusts the corrupt device. Some protocol designers might dismiss the problem at this point, thinking that it is outside their threat model. One counterargument is that the corrupt device is not the one that ends up being paired, and thus the honest device C also suffers. Another is that the Internet of Things will be full of corrupt insiders, just like the regular Internet. Also, we should protect the users from their own mistakes whenever possible.

The practical impact of misbinding attacks is somewhat difficult to grasp. It has been demonstrated with the help of two example scenarios, one presented by Diffie et al.~in the original STS paper and the other by Krawczyk in a lecture:
\begin{itemize}
\item A connects to bank C, over a supposedly secure session, to deposit an electronic coin. Since B mounted a misbinding attack, bank C thinks the coin was deposited by B. 
\item B and C are fighter jets, and A is their commander. B has been compromised by the enemy. A tells B to self-destruct, but because B mounted a misbinding attack, the command goes to C. 
\end{itemize}

The banking scenario does not seem to have obvious equivalents in the world of physical devices. The fighter jets, on the other hand, are devices, and we can construct a related IoT example:
\begin{itemize}
\item B and C are IoT devices, and A is the user's computer. B has been infected by malware. User wants to connect A to B and wipe B's memory. Because B mounts a misbinding attack, the user wipes C instead. 
\end{itemize}
Note that all these scenarios require some prior relation between the endpoints, and the misbinding attack leads to a failure of correspondence between that prior relation and the newly established connection. In pairing and bootstrapping, there often is no such common history. Either one of the endpoints is a new, fresh device, or the history is not significant because the endpoints have no secure way of knowing that they have reconnected to the same peer. This may be one reason why the practical impact of misbinding for IoT devices remains somewhat elusive.  

We also need to compare misbinding to alternative attacks. In Figure~\ref{fig:misbinding-pairing}(b), the attacker in device B can achieve almost the same results by accepting the connection from A, establishing another connection to C, and then forwarding the application-layer messages between A and C. The main difference between misbinding and such relaying of communication is that the misbinding attacker can remove itself from the communication chain after the pairing. Thus, the continuation of the attack does not depend on the compromised device B being online or within radio range. Furthermore, the user is in physical control of the compromised device B but not of C. If the user disables device B, e.g.~by disconnecting it from the network or even by physically destroying it, device C and is connection with A will nevertheless persists --- unknown to the user.

Since the design of STS and IKE, there has been consensus among security protocol designers that misbinding vulnerabilities are not acceptable in authenticated key-exchange protocols for computer networks and for the Internet. In device pairing, there is no similar consensus, and the attack has been mostly ignored with the exception of the trusted-computing community. We do not expect this paper to stop people from using protocols like Bluetooth SSP. The misbinding attacks and impact scenarios are relatively marginal compared to the advantage of encrypting wireless communication and having basic authentication in place, and the value of these is not nullified by misbinding. The attacks should, however, not be ignored because they are so widely applicable to device-pairing and IoT bootstrapping protocols. Our message is that protocol and system designers should understand the misbinding vulnerability for physical devices, keep eyes open for unexpected consequences in new situations, and make a balanced judgment about whether additional countermeasures are needed.

%% file: conclusion.tex

\section{Conclusion}
\label{sec:conclusion}

We studied identity-misbinding (or unknown-key-share) attacks in device pairing protocols where the devices are identified by physical access rather than cryptographic credentials. We showed that Bluetooth and other similar device-pairing protocols are vulnerable to this attack regardless of their cryptographic details. The same vulnerability also exists in protocols for security-bootstrapping IoT devices. We confirmed the attacks by implementing them. Formal modelling allowed us to discuss the precise definition of misbinding, which led to the discovery of a new attack variant, double misbinding. We also discussed potential mitigation mechanisms, arguing in favor of solutions based on asset tracking. While the vulnerability to identity misbinding does not make the existing device pairing protocols completely insecure, it is a threat that needs to be fully understood also in device pairing, and this paper is a step towards that goal.

%% file: acknowledgment.tex
\section{Acknowledgments}
\label{sec:acknowledgments}

We would like to thank Eric Rescorla for his inspiring comments on EAP-NOOB and Kaisa Nyberg for insightful discussion on Bluetooth SSP. This work was supported by Academy of Finland (grant number 296693).

%% file: main.bbl

\begin{thebibliography}{46}


\ifx \showCODEN    \undefined \def \showCODEN     #1{\unskip}     \fi
\ifx \showDOI      \undefined \def \showDOI       #1{#1}\fi
\ifx \showISBNx    \undefined \def \showISBNx     #1{\unskip}     \fi
\ifx \showISBNxiii \undefined \def \showISBNxiii  #1{\unskip}     \fi
\ifx \showISSN     \undefined \def \showISSN      #1{\unskip}     \fi
\ifx \showLCCN     \undefined \def \showLCCN      #1{\unskip}     \fi
\ifx \shownote     \undefined \def \shownote      #1{#1}          \fi
\ifx \showarticletitle \undefined \def \showarticletitle #1{#1}   \fi
\ifx \showURL      \undefined \def \showURL       {\relax}        \fi
\providecommand\bibfield[2]{#2}
\providecommand\bibinfo[2]{#2}
\providecommand\natexlab[1]{#1}
\providecommand\showeprint[2][]{arXiv:#2}

\bibitem[\protect\citeauthoryear{11889-1:2015}{11889-1:2015}{2015}]%
        {tpm2}
\bibfield{author}{\bibinfo{person}{ISO/IEC 11889-1:2015}.}
  \bibinfo{year}{2015}\natexlab{}.
\newblock \bibinfo{booktitle}{\emph{Information technology -- Trusted platform
  module library -- Part 1: Architecture}}.
\newblock \bibinfo{type}{Standard}. \bibinfo{institution}{International
  Organization for Standardization}.
\newblock


\bibitem[\protect\citeauthoryear{Abadi and Needham}{Abadi and Needham}{1996}]%
        {abadi1996prudent}
\bibfield{author}{\bibinfo{person}{Martin Abadi} {and} \bibinfo{person}{Roger
  Needham}.} \bibinfo{year}{1996}\natexlab{}.
\newblock \showarticletitle{Prudent Engineering Practice for Cryptographic
  Protocols}.
\newblock \bibinfo{journal}{\emph{Transactions on Software Engineering}}
  \bibinfo{volume}{22}, \bibinfo{number}{1} (\bibinfo{year}{1996}),
  \bibinfo{pages}{6--15}.
\newblock


\bibitem[\protect\citeauthoryear{Aboba, Blunk, Vollbrecht, Carlson, and
  Levkowetz}{Aboba et~al\mbox{.}}{2004}]%
        {rfc3748}
\bibfield{author}{\bibinfo{person}{Bernard Aboba}, \bibinfo{person}{Larry~J.
  Blunk}, \bibinfo{person}{John~R. Vollbrecht}, \bibinfo{person}{James
  Carlson}, {and} \bibinfo{person}{Henrik Levkowetz}.}
  \bibinfo{year}{2004}\natexlab{}.
\newblock \bibinfo{title}{Extensible Authentication Protocol ({EAP})}.
\newblock \bibinfo{howpublished}{\url{http://tools.ietf.org/rfc/rfc3748.txt}}.
\newblock
\newblock
\shownote{RFC 3748.}


\bibitem[\protect\citeauthoryear{Alliance}{Alliance}{2016}]%
        {wifi-direct}
\bibfield{author}{\bibinfo{person}{Wi-Fi Alliance}.}
  \bibinfo{year}{2016}\natexlab{}.
\newblock \bibinfo{booktitle}{\emph{{Wi-Fi} Peer-to-Peer ({P2P}) Technical
  Specification, v. 1.7}}.
\newblock \bibinfo{type}{{T}echnical {R}eport}. \bibinfo{institution}{Wi-Fi
  Alliance}.
\newblock


\bibitem[\protect\citeauthoryear{Alliance}{Alliance}{2018}]%
        {DPP}
\bibfield{author}{\bibinfo{person}{Wi-Fi Alliance}.}
  \bibinfo{year}{2018}\natexlab{}.
\newblock \bibinfo{booktitle}{\emph{Device Provisioning Protocol Specification
  Version 1.0}}.
\newblock \bibinfo{type}{{T}echnical {R}eport}. \bibinfo{institution}{Wi-Fi
  Alliance}.
\newblock


\bibitem[\protect\citeauthoryear{Alliance}{Alliance}{2012}]%
        {alliance2012zigbee}
\bibfield{author}{\bibinfo{person}{ZigBee Alliance}.}
  \bibinfo{year}{2012}\natexlab{}.
\newblock \bibinfo{booktitle}{\emph{{ZigBee} Specification}}.
\newblock \bibinfo{type}{ZigBee Alliance Document} 053474r20.
  \bibinfo{institution}{ZigBee Alliance}.
\newblock


\bibitem[\protect\citeauthoryear{Armando, Basin, Boichut, Chevalier, Compagna,
  Cu{\'e}llar, Drielsma, H{\'e}am, Kouchnarenko, Mantovani,
  et~al\mbox{.}}{Armando et~al\mbox{.}}{2005}]%
        {armando2005avispa}
\bibfield{author}{\bibinfo{person}{Alessandro Armando}, \bibinfo{person}{David
  Basin}, \bibinfo{person}{Yohan Boichut}, \bibinfo{person}{Yannick Chevalier},
  \bibinfo{person}{Luca Compagna}, \bibinfo{person}{Jorge Cu{\'e}llar},
  \bibinfo{person}{P.~Hankes Drielsma}, \bibinfo{person}{Pierre-Cyrille
  H{\'e}am}, \bibinfo{person}{Olga Kouchnarenko}, \bibinfo{person}{Jacopo
  Mantovani}, {et~al\mbox{.}}} \bibinfo{year}{2005}\natexlab{}.
\newblock \showarticletitle{The {AVISPA} Tool for the Automated Validation of
  Internet Security Protocols and Applications}. In
  \bibinfo{booktitle}{\emph{Proceedings of the International conference on
  computer aided verification}}. \bibinfo{publisher}{Springer Berlin
  Heidelberg}, \bibinfo{address}{Berlin, Heidelberg},
  \bibinfo{pages}{281--285}.
\newblock


\bibitem[\protect\citeauthoryear{Aura and Sethi}{Aura and Sethi}{2019}]%
        {aura-eap-noob-05}
\bibfield{author}{\bibinfo{person}{Tuomas Aura} {and} \bibinfo{person}{Mohit
  Sethi}.} \bibinfo{year}{2019}\natexlab{}.
\newblock \bibinfo{booktitle}{\emph{Nimble out-of-band authentication for {EAP
  (EAP-NOOB)}}}.
\newblock \bibinfo{type}{Internet-Draft} draft-aura-eap-noob-05.
  \bibinfo{institution}{Internet Engineering Task Force}.
\newblock


\bibitem[\protect\citeauthoryear{Blake-Wilson and Menezes}{Blake-Wilson and
  Menezes}{1999}]%
        {blake1999unknown}
\bibfield{author}{\bibinfo{person}{Simon Blake-Wilson} {and}
  \bibinfo{person}{Alfred Menezes}.} \bibinfo{year}{1999}\natexlab{}.
\newblock \showarticletitle{Unknown Key-Share Attacks on the Station-to-Station
  {(STS)} Protocol}. In \bibinfo{booktitle}{\emph{Proceedings of the
  International Workshop on Public Key Cryptography}}.
  \bibinfo{publisher}{Springer-Verlag}, \bibinfo{address}{Berlin, Heidelberg},
  \bibinfo{pages}{154--170}.
\newblock


\bibitem[\protect\citeauthoryear{Blanchet}{Blanchet}{2001}]%
        {blanchet2001efficient}
\bibfield{author}{\bibinfo{person}{Bruno Blanchet}.}
  \bibinfo{year}{2001}\natexlab{}.
\newblock \showarticletitle{An Efficient Cryptographic Protocol Verifier Based
  on {Prolog} Rules}. In \bibinfo{booktitle}{\emph{Proceedings of the 14th
  Computer Security Foundations Workshop}}. \bibinfo{publisher}{IEEE},
  \bibinfo{pages}{82--96}.
\newblock


\bibitem[\protect\citeauthoryear{Blanchet, Smyth, Cheval, and
  Sylvestre}{Blanchet et~al\mbox{.}}{2018}]%
        {blanchet2018proverif}
\bibfield{author}{\bibinfo{person}{Bruno Blanchet}, \bibinfo{person}{Ben
  Smyth}, \bibinfo{person}{Vincent Cheval}, {and} \bibinfo{person}{Marc
  Sylvestre}.} \bibinfo{year}{2018}\natexlab{}.
\newblock \bibinfo{booktitle}{\emph{{ProVerif} 2.00: Automatic Cryptographic
  Protocol Verifier, User Manual and Tutorial}}.
\newblock INRIA.
\newblock


\bibitem[\protect\citeauthoryear{Carlos}{Carlos}{2014}]%
        {carlos2014towards}
\bibfield{author}{\bibinfo{person}{Marcelo~Carlomagno Carlos}.}
  \bibinfo{year}{2014}\natexlab{}.
\newblock \emph{\bibinfo{title}{Towards a Multidisciplinary Framework for the
  Design and Analysis of Security Ceremonies}}.
\newblock \bibinfo{thesistype}{Ph.D. Dissertation}. \bibinfo{school}{Royal
  Holloway, University of London}.
\newblock


\bibitem[\protect\citeauthoryear{Carlos, Martina, Price, and
  Cust{\'o}dio}{Carlos et~al\mbox{.}}{2013}]%
        {carlos2013updated}
\bibfield{author}{\bibinfo{person}{Marcelo~Carlomagno Carlos},
  \bibinfo{person}{Jean~Everson Martina}, \bibinfo{person}{Geraint Price},
  {and} \bibinfo{person}{Ricardo~Felipe Cust{\'o}dio}.}
  \bibinfo{year}{2013}\natexlab{}.
\newblock \showarticletitle{An Updated Threat Model for Security Ceremonies}.
  In \bibinfo{booktitle}{\emph{Proceedings of the 28th Annual ACM Symposium on
  Applied Computing}}. \bibinfo{publisher}{ACM}, \bibinfo{address}{New York,
  NY, USA}, \bibinfo{pages}{1836--1843}.
\newblock


\bibitem[\protect\citeauthoryear{Chang and Shmatikov}{Chang and
  Shmatikov}{2007}]%
        {chang2007formal}
\bibfield{author}{\bibinfo{person}{Richard Chang} {and} \bibinfo{person}{Vitaly
  Shmatikov}.} \bibinfo{year}{2007}\natexlab{}.
\newblock \showarticletitle{Formal Analysis of Authentication in {B}luetooth
  Device Pairing}. In \bibinfo{booktitle}{\emph{Proceedings of the Joint
  Workshop on Foundations of Computer Security and Automated Reasoning for
  Security Protocol Analysis}}.
\newblock


\bibitem[\protect\citeauthoryear{Diffie, Van~Oorschot, and Wiener}{Diffie
  et~al\mbox{.}}{1992}]%
        {diffie1992authentication}
\bibfield{author}{\bibinfo{person}{Whitfield Diffie}, \bibinfo{person}{Paul~C.
  Van~Oorschot}, {and} \bibinfo{person}{Michael~J. Wiener}.}
  \bibinfo{year}{1992}\natexlab{}.
\newblock \showarticletitle{Authentication and Authenticated Key Exchanges}.
\newblock \bibinfo{journal}{\emph{Designs, Codes and cryptography}}
  \bibinfo{volume}{2}, \bibinfo{number}{2} (\bibinfo{year}{1992}),
  \bibinfo{pages}{107--125}.
\newblock


\bibitem[\protect\citeauthoryear{Digicert}{Digicert}{2019}]%
        {digicert}
\bibfield{author}{\bibinfo{person}{Digicert}.} \bibinfo{year}{2019}\natexlab{}.
\newblock \bibinfo{title}{Device Certificates}.
\newblock
  \bibinfo{howpublished}{\url{https://www.digicert.com/device-certificates/}}.
\newblock
\newblock
\shownote{Accessed: 11.5.2019.}


\bibitem[\protect\citeauthoryear{Dill}{Dill}{1996}]%
        {dill1996mur}
\bibfield{author}{\bibinfo{person}{David~L. Dill}.}
  \bibinfo{year}{1996}\natexlab{}.
\newblock \showarticletitle{The {Murphi} Verification System}. In
  \bibinfo{booktitle}{\emph{Proceedings of the International Conference on
  Computer Aided Verification}}. \bibinfo{publisher}{Springer-Verlag},
  \bibinfo{address}{London, UK, UK}, \bibinfo{pages}{390--393}.
\newblock


\bibitem[\protect\citeauthoryear{Ding and Tsudik}{Ding and Tsudik}{2018}]%
        {ding2018initializing}
\bibfield{author}{\bibinfo{person}{Xuhua Ding} {and} \bibinfo{person}{Gene
  Tsudik}.} \bibinfo{year}{2018}\natexlab{}.
\newblock \showarticletitle{Initializing trust in smart devices via presence
  attestation}.
\newblock \bibinfo{journal}{\emph{Computer Communications}}
  \bibinfo{volume}{131} (\bibinfo{year}{2018}), \bibinfo{pages}{35--38}.
\newblock


\bibitem[\protect\citeauthoryear{Dolev and Yao}{Dolev and Yao}{1983}]%
        {dolev1983security}
\bibfield{author}{\bibinfo{person}{Danny Dolev} {and} \bibinfo{person}{Andrew
  Yao}.} \bibinfo{year}{1983}\natexlab{}.
\newblock \showarticletitle{On the Security of Public Key Protocols}.
\newblock \bibinfo{journal}{\emph{Transactions on Information Theory}}
  \bibinfo{volume}{29}, \bibinfo{number}{2} (\bibinfo{year}{1983}),
  \bibinfo{pages}{198--208}.
\newblock


\bibitem[\protect\citeauthoryear{Ellison}{Ellison}{2007}]%
        {ellison2007ceremony}
\bibfield{author}{\bibinfo{person}{Carl~M. Ellison}.}
  \bibinfo{year}{2007}\natexlab{}.
\newblock \showarticletitle{Ceremony Design and Analysis}.
\newblock \bibinfo{journal}{\emph{IACR Cryptology ePrint Archive}}
  (\bibinfo{year}{2007}).
\newblock


\bibitem[\protect\citeauthoryear{Fink, Sherman, Mitchell, and Challener}{Fink
  et~al\mbox{.}}{2011}]%
        {fink2011catching}
\bibfield{author}{\bibinfo{person}{Russell~A. Fink}, \bibinfo{person}{Alan~T.
  Sherman}, \bibinfo{person}{Alexander~O. Mitchell}, {and}
  \bibinfo{person}{David~C. Challener}.} \bibinfo{year}{2011}\natexlab{}.
\newblock \showarticletitle{Catching the Cuckoo: Verifying TPM Proximity Using
  a Quote Timing Side-channel}. In \bibinfo{booktitle}{\emph{Proceedings of the
  International Conference on Trust and Trustworthy Computing}}.
  \bibinfo{publisher}{Springer-Verlag}, \bibinfo{address}{Berlin, Heidelberg},
  \bibinfo{pages}{294--301}.
\newblock


\bibitem[\protect\citeauthoryear{Gajbhiye, Sharma, Karmkar, and
  Sharma}{Gajbhiye et~al\mbox{.}}{2016}]%
        {gajbhiye2016design}
\bibfield{author}{\bibinfo{person}{Samta Gajbhiye}, \bibinfo{person}{Monisha
  Sharma}, \bibinfo{person}{Sanjeev Karmkar}, {and} \bibinfo{person}{Sanjay
  Sharma}.} \bibinfo{year}{2016}\natexlab{}.
\newblock \showarticletitle{Design, Implementation and Security Analysis of
  {B}luetooth Pairing Protocol in {NS2}}. In
  \bibinfo{booktitle}{\emph{Proceedings of the International Conference on
  Advances in Computing, Communications and Informatics (ICACCI)}}.
  \bibinfo{publisher}{IEEE}, \bibinfo{pages}{1711--1717}.
\newblock


\bibitem[\protect\citeauthoryear{Goldman, Perez, and Sailer}{Goldman
  et~al\mbox{.}}{2006}]%
        {goldman2006linking}
\bibfield{author}{\bibinfo{person}{Kenneth Goldman}, \bibinfo{person}{Ronald
  Perez}, {and} \bibinfo{person}{Reiner Sailer}.}
  \bibinfo{year}{2006}\natexlab{}.
\newblock \showarticletitle{Linking Remote Attestation to Secure Tunnel
  Endpoints}. In \bibinfo{booktitle}{\emph{Proceedings of the first workshop on
  Scalable trusted computing}}. \bibinfo{publisher}{ACM}, \bibinfo{address}{New
  York, NY, USA}, \bibinfo{pages}{21--24}.
\newblock


\bibitem[\protect\citeauthoryear{Haataja and Toivanen}{Haataja and
  Toivanen}{2010}]%
        {haataja2010two}
\bibfield{author}{\bibinfo{person}{Keijo Haataja} {and} \bibinfo{person}{Pekka
  Toivanen}.} \bibinfo{year}{2010}\natexlab{}.
\newblock \showarticletitle{Two Practical Man-in-the-middle Attacks on
  Bluetooth Secure Simple Pairing and Countermeasures}.
\newblock \bibinfo{journal}{\emph{Transactions on Wireless Communications}}
  \bibinfo{volume}{9}, \bibinfo{number}{1} (\bibinfo{year}{2010}),
  \bibinfo{pages}{384--392}.
\newblock


\bibitem[\protect\citeauthoryear{Hancke and Kuhn}{Hancke and Kuhn}{2005}]%
        {hancke2005rfid}
\bibfield{author}{\bibinfo{person}{Gerhard~P. Hancke} {and}
  \bibinfo{person}{Markus~G. Kuhn}.} \bibinfo{year}{2005}\natexlab{}.
\newblock \showarticletitle{An {RFID} Distance Bounding Protocol}. In
  \bibinfo{booktitle}{\emph{Security and Privacy for Emerging Areas in
  Communications Networks, 2005. SecureComm 2005. First International
  Conference on}}. \bibinfo{publisher}{IEEE Computer Society},
  \bibinfo{address}{Washington, DC, USA}, \bibinfo{pages}{67--73}.
\newblock


\bibitem[\protect\citeauthoryear{Hassan, Bibon, Hossain, and
  Atiquzzaman}{Hassan et~al\mbox{.}}{2018}]%
        {hassan2018security}
\bibfield{author}{\bibinfo{person}{Shaikh~Shahriar Hassan},
  \bibinfo{person}{Soumik~Das Bibon}, \bibinfo{person}{Md~Shohrab Hossain},
  {and} \bibinfo{person}{Mohammed Atiquzzaman}.}
  \bibinfo{year}{2018}\natexlab{}.
\newblock \showarticletitle{Security Threats in {B}luetooth Technology}.
\newblock \bibinfo{journal}{\emph{Computers \& Security}}  \bibinfo{volume}{74}
  (\bibinfo{year}{2018}), \bibinfo{pages}{308--322}.
\newblock


\bibitem[\protect\citeauthoryear{IEEE}{IEEE}{2009}]%
        {5367679}
\bibfield{author}{\bibinfo{person}{IEEE}.} \bibinfo{year}{2009}\natexlab{}.
\newblock \bibinfo{booktitle}{\emph{{IEEE} Standard for Local and metropolitan
  area networks - Secure Device Identity}}.
\newblock \bibinfo{type}{{T}echnical {R}eport}. \bibinfo{institution}{IEEE}.
  \bibinfo{pages}{1--77} pages.
\newblock
\urldef\tempurl%
\url{https://doi.org/10.1109/IEEESTD.2009.5367679}
\showDOI{\tempurl}


\bibitem[\protect\citeauthoryear{IEEE}{IEEE}{2016}]%
        {7786995}
\bibfield{author}{\bibinfo{person}{IEEE}.} \bibinfo{year}{2016}\natexlab{}.
\newblock \bibinfo{booktitle}{\emph{{IEEE} Standard for Information
  technology--Telecommunications and information exchange between systems Local
  and metropolitan area networks--Specific requirements - Part 11: Wireless
  {LAN} Medium Access Control ({MAC}) and Physical Layer ({PHY})
  Specifications}}.
\newblock \bibinfo{type}{{T}echnical {R}eport}. \bibinfo{institution}{IEEE}.
  \bibinfo{pages}{1--3534} pages.
\newblock


\bibitem[\protect\citeauthoryear{Jia and Hsu}{Jia and Hsu}{2013}]%
        {jiaformal}
\bibfield{author}{\bibinfo{person}{David Jia} {and} \bibinfo{person}{Richard
  Hsu}.} \bibinfo{year}{2013}\natexlab{}.
\newblock \bibinfo{title}{Formal Modeling and Analysis of {B}luetooth 4.0
  Pairing Protocol}.
\newblock
\newblock


\bibitem[\protect\citeauthoryear{Kainda, Flechais, and Roscoe}{Kainda
  et~al\mbox{.}}{2009}]%
        {kainda2009usability}
\bibfield{author}{\bibinfo{person}{Ronald Kainda}, \bibinfo{person}{Ivan
  Flechais}, {and} \bibinfo{person}{A.~W. Roscoe}.}
  \bibinfo{year}{2009}\natexlab{}.
\newblock \showarticletitle{Usability and Security of Out-of-band Channels in
  Secure Device Pairing Protocols}. In \bibinfo{booktitle}{\emph{Proceedings of
  the 5th Symposium on Usable Privacy and Security}}. \bibinfo{publisher}{ACM},
  \bibinfo{address}{New York, NY, USA}, \bibinfo{pages}{11:1--11:12}.
\newblock


\bibitem[\protect\citeauthoryear{Kaufman, Hoffman, Nir, Eronen, and
  Kivinen}{Kaufman et~al\mbox{.}}{2014}]%
        {rfc7296}
\bibfield{author}{\bibinfo{person}{Charlie Kaufman}, \bibinfo{person}{Paul~E.
  Hoffman}, \bibinfo{person}{Yoav Nir}, \bibinfo{person}{Pasi Eronen}, {and}
  \bibinfo{person}{Tero Kivinen}.} \bibinfo{year}{2014}\natexlab{}.
\newblock \bibinfo{title}{Internet Key Exchange Protocol Version 2 ({IKEv2})}.
\newblock \bibinfo{howpublished}{\url{http://tools.ietf.org/rfc/rfc7296.txt}}.
\newblock
\newblock
\shownote{RFC 7296.}


\bibitem[\protect\citeauthoryear{Kravitz and Cooper}{Kravitz and
  Cooper}{2017}]%
        {kravitz2017securing}
\bibfield{author}{\bibinfo{person}{David~W. Kravitz} {and}
  \bibinfo{person}{Jason Cooper}.} \bibinfo{year}{2017}\natexlab{}.
\newblock \showarticletitle{Securing User Identity and Transactions
  Symbiotically: {IoT} Meets Blockchain}. In
  \bibinfo{booktitle}{\emph{Proceedings of the Global Internet of Things Summit
  (GIoTS)}}. \bibinfo{publisher}{IEEE}, \bibinfo{pages}{1--6}.
\newblock


\bibitem[\protect\citeauthoryear{Krawczyk}{Krawczyk}{2003}]%
        {krawczyk2003sigma}
\bibfield{author}{\bibinfo{person}{Hugo Krawczyk}.}
  \bibinfo{year}{2003}\natexlab{}.
\newblock \showarticletitle{{SIGMA: The `SIGn-and-MAc'} Approach to
  Authenticated {Diffie-Hellman} and Its Use in the {IKE} Protocols}. In
  \bibinfo{booktitle}{\emph{Annual International Cryptology Conference}}.
  \bibinfo{publisher}{Springer Berlin Heidelberg}, \bibinfo{address}{Berlin,
  Heidelberg}, \bibinfo{pages}{400--425}.
\newblock


\bibitem[\protect\citeauthoryear{Kuo, Luk, Negi, and Perrig}{Kuo
  et~al\mbox{.}}{2007}]%
        {kuo2007message}
\bibfield{author}{\bibinfo{person}{Cynthia Kuo}, \bibinfo{person}{Mark Luk},
  \bibinfo{person}{Rohit Negi}, {and} \bibinfo{person}{Adrian Perrig}.}
  \bibinfo{year}{2007}\natexlab{}.
\newblock \showarticletitle{Message-in-a-bottle: User-friendly and Secure Key
  Deployment for Sensor Nodes}. In \bibinfo{booktitle}{\emph{Proceedings of the
  5th international conference on Embedded networked sensor systems}}.
  \bibinfo{publisher}{ACM}, \bibinfo{address}{New York, NY, USA},
  \bibinfo{pages}{233--246}.
\newblock


\bibitem[\protect\citeauthoryear{Levi, {\c{C}}etinta{\c{s}}, Aydos, Ko{\c{c}},
  and {\c{C}}a{\u{g}}layan}{Levi et~al\mbox{.}}{2004}]%
        {levi2004relay}
\bibfield{author}{\bibinfo{person}{Albert Levi}, \bibinfo{person}{Erhan
  {\c{C}}etinta{\c{s}}}, \bibinfo{person}{Murat Aydos},
  \bibinfo{person}{{\c{C}}etin~Kaya Ko{\c{c}}}, {and} \bibinfo{person}{M.~Ufuk
  {\c{C}}a{\u{g}}layan}.} \bibinfo{year}{2004}\natexlab{}.
\newblock \showarticletitle{Relay Attacks on {B}luetooth Authentication and
  Solutions}. In \bibinfo{booktitle}{\emph{International Symposium on Computer
  and Information Sciences}}. \bibinfo{publisher}{Springer Berlin Heidelberg},
  \bibinfo{address}{Berlin, Heidelberg}, \bibinfo{pages}{278--288}.
\newblock


\bibitem[\protect\citeauthoryear{Lowe}{Lowe}{1995}]%
        {lowe1995attack}
\bibfield{author}{\bibinfo{person}{Gavin Lowe}.}
  \bibinfo{year}{1995}\natexlab{}.
\newblock \showarticletitle{An Attack on the {Needham-Schroeder} Public-Key
  Authentication Protocol}.
\newblock \bibinfo{journal}{\emph{Inform. Process. Lett.}}
  \bibinfo{volume}{56}, \bibinfo{number}{3} (\bibinfo{year}{1995}),
  \bibinfo{pages}{131--133}.
\newblock


\bibitem[\protect\citeauthoryear{Lowe}{Lowe}{1997}]%
        {lowe1997hierarchy}
\bibfield{author}{\bibinfo{person}{Gavin Lowe}.}
  \bibinfo{year}{1997}\natexlab{}.
\newblock \showarticletitle{A Hierarchy of Authentication Specifications}. In
  \bibinfo{booktitle}{\emph{Proceedings of the 10th Computer security
  foundations workshop}}. \bibinfo{publisher}{IEEE Computer Society},
  \bibinfo{address}{Washington, DC, USA}, \bibinfo{pages}{31--43}.
\newblock


\bibitem[\protect\citeauthoryear{Naveed, Zhou, Demetriou, Wang, and
  Gunter}{Naveed et~al\mbox{.}}{2014}]%
        {naveed2014inside}
\bibfield{author}{\bibinfo{person}{Muhammad Naveed}, \bibinfo{person}{Xiaoyong
  Zhou}, \bibinfo{person}{Soteris Demetriou}, \bibinfo{person}{XiaoFeng Wang},
  {and} \bibinfo{person}{Carl~A. Gunter}.} \bibinfo{year}{2014}\natexlab{}.
\newblock \showarticletitle{Inside Job: Understanding and Mitigating the Threat
  of External Device Mis-Binding on {A}ndroid}. In
  \bibinfo{booktitle}{\emph{Proceedings of the Network and Distributed System
  Security Symposium (NDSS)}}.
\newblock


\bibitem[\protect\citeauthoryear{Nuss, Puchta, and Kunz}{Nuss
  et~al\mbox{.}}{2018}]%
        {nuss2018towards}
\bibfield{author}{\bibinfo{person}{Martin Nuss}, \bibinfo{person}{Alexander
  Puchta}, {and} \bibinfo{person}{Michael Kunz}.}
  \bibinfo{year}{2018}\natexlab{}.
\newblock \showarticletitle{Towards Blockchain-Based Identity and Access
  Management for {Internet of Things} in Enterprises}. In
  \bibinfo{booktitle}{\emph{Proceedings of the International Conference on
  Trust and Privacy in Digital Business}}. \bibinfo{publisher}{Springer
  International Publishing}, \bibinfo{address}{Cham},
  \bibinfo{pages}{167--181}.
\newblock


\bibitem[\protect\citeauthoryear{Parno, McCune, and Perrig}{Parno
  et~al\mbox{.}}{2011}]%
        {parno2011bootstrapping}
\bibfield{author}{\bibinfo{person}{Bryan Parno}, \bibinfo{person}{Jonathan~M
  McCune}, {and} \bibinfo{person}{Adrian Perrig}.}
  \bibinfo{year}{2011}\natexlab{}.
\newblock \bibinfo{booktitle}{\emph{Bootstrapping Trust in Modern Computers}}.
\newblock \bibinfo{publisher}{Springer Science \& Business Media}.
\newblock


\bibitem[\protect\citeauthoryear{Rasmussen and Capkun}{Rasmussen and
  Capkun}{2010}]%
        {rasmussen2010realization}
\bibfield{author}{\bibinfo{person}{Kasper~Bonne Rasmussen} {and}
  \bibinfo{person}{Srdjan Capkun}.} \bibinfo{year}{2010}\natexlab{}.
\newblock \showarticletitle{Realization of {RF} Distance Bounding}. In
  \bibinfo{booktitle}{\emph{Proceedings of the USENIX Security Symposium}}.
  \bibinfo{publisher}{USENIX Association}, \bibinfo{address}{Berkeley, CA,
  USA}, \bibinfo{pages}{389--402}.
\newblock


\bibitem[\protect\citeauthoryear{Saxena, Ekberg, Kostiainen, and Asokan}{Saxena
  et~al\mbox{.}}{2006}]%
        {saxena2006secure}
\bibfield{author}{\bibinfo{person}{Nitesh Saxena}, \bibinfo{person}{Jan-Erik
  Ekberg}, \bibinfo{person}{Kari Kostiainen}, {and} \bibinfo{person}{N.
  Asokan}.} \bibinfo{year}{2006}\natexlab{}.
\newblock \showarticletitle{Secure Device Pairing based on a Visual Channel}.
  In \bibinfo{booktitle}{\emph{Proceedings of the Symposium on Security and
  Privacy}}. \bibinfo{publisher}{IEEE}.
\newblock


\bibitem[\protect\citeauthoryear{SIG}{SIG}{2016}]%
        {bluetooth2016}
\bibfield{author}{\bibinfo{person}{Bluetooth SIG}.}
  \bibinfo{year}{2016}\natexlab{}.
\newblock \bibinfo{booktitle}{\emph{{B}luetooth Specification Version 5.0}}.
\newblock \bibinfo{type}{Core Specification}. \bibinfo{institution}{Bluetooth
  SIG}.
\newblock
\newblock
\shownote{\url{https://www.bluetooth.com/specifications/bluetooth-core-specification}.}


\bibitem[\protect\citeauthoryear{Suomalainen, Valkonen, and Asokan}{Suomalainen
  et~al\mbox{.}}{2007}]%
        {suomalainen2007security}
\bibfield{author}{\bibinfo{person}{Jani Suomalainen}, \bibinfo{person}{Jukka
  Valkonen}, {and} \bibinfo{person}{N. Asokan}.}
  \bibinfo{year}{2007}\natexlab{}.
\newblock \showarticletitle{Security Associations in Personal Networks: A
  Comparative Analysis}. In \bibinfo{booktitle}{\emph{Proceedings of the
  European Workshop on Security in Ad-hoc and Sensor Networks}}.
  \bibinfo{publisher}{Springer Berlin Heidelberg}, \bibinfo{address}{Berlin,
  Heidelberg}, \bibinfo{pages}{43--57}.
\newblock


\bibitem[\protect\citeauthoryear{Woo and Lam}{Woo and Lam}{1993}]%
        {woo1993semantic}
\bibfield{author}{\bibinfo{person}{Thomas~Y.C. Woo} {and}
  \bibinfo{person}{Simon~S. Lam}.} \bibinfo{year}{1993}\natexlab{}.
\newblock \showarticletitle{A Semantic Model for Authentication Protocols}. In
  \bibinfo{booktitle}{\emph{Proceedings of the IEEE Computer Society Symposium
  on Research in Security and Privacy}}. \bibinfo{publisher}{IEEE Computer
  Society}, \bibinfo{address}{Washington, DC, USA}, \bibinfo{pages}{178--194}.
\newblock


\bibitem[\protect\citeauthoryear{Zhang, Ding, Tsudik, Cui, and Li}{Zhang
  et~al\mbox{.}}{2017}]%
        {zhang2017presence}
\bibfield{author}{\bibinfo{person}{Zhangkai Zhang}, \bibinfo{person}{Xuhua
  Ding}, \bibinfo{person}{Gene Tsudik}, \bibinfo{person}{Jinhua Cui}, {and}
  \bibinfo{person}{Zhoujun Li}.} \bibinfo{year}{2017}\natexlab{}.
\newblock \showarticletitle{Presence Attestation: The Missing Link in Dynamic
  Trust Bootstrapping}. In \bibinfo{booktitle}{\emph{Proceedings of the ACM
  SIGSAC Conference on Computer and Communications Security}}.
  \bibinfo{publisher}{ACM}, \bibinfo{address}{New York, NY, USA},
  \bibinfo{pages}{89--102}.
\newblock


\end{thebibliography}
